\documentclass[%
 reprint,
 amsmath,amssymb,
 aps,
]{revtex4-2}
\usepackage[utf8]{inputenc}
\usepackage[T1]{fontenc}
\usepackage{color}
\usepackage{array}
\usepackage{float}
\usepackage{booktabs}
\usepackage{multirow}
\usepackage{amsmath}
\usepackage{amssymb}
\usepackage{stackrel}
\usepackage{graphicx}
\usepackage{subcaption}
\usepackage{tabularx}
\usepackage{xr}

\makeatletter
\usepackage[format=plain,justification=justified,singlelinecheck=false]{caption}
\usepackage{ragged2e} 

\newcommand{\lyxmathsym}[1]{\ifmmode\begingroup\def\b@ld{bold}
  \text{\ifx\math@version\b@ld\bfseries\fi#1}\endgroup\else#1\fi}

\providecommand{\tabularnewline}{\\}

\makeatother

\usepackage{babel}

\begin{document}

\title{Accelerated Prediction of Temperature-Dependent Lattice Thermal Conductivity via Ensembled Machine Learning Models}
\author{Piyush Paliwal$^1$}
\author{Aftab Alam$^{1,2}$}
\email{aftab@iitb.ac.in}
\affiliation{$^1$Centre for Machine Intelligence and Data Science (CMInDS), Indian Institute of Technology Bombay, Powai, Mumbai 400076, India}
\affiliation{$^2$Department of Physics, Indian Institute of Technology Bombay, Powai, Mumbai 400076, India}

\begin{abstract}
Lattice thermal conductivity ($\kappa_L$) is a key physical property governing heat transport in solids, with direct relevance to thermoelectrics, thermal barrier coatings, and heat management applications. However, while experimental determination of $\kappa_L$ is challenging, its theoretical calculation via ab initio methods particularly using density functional theory (DFT) is computationally intensive, often more demanding than electronic transport calculations by an order of magnitude. In this work, we present a machine learning (ML) approach to predict $\kappa_L$ with DFT-level accuracy over a wide temperature range (100-1000 K). Among various models trained on DFT-calculated data \textcolor{black}{obtained} from literature, the Extra Trees Regressor (ETR) yielded the best performance on log-scaled $\kappa_L$, achieving an average $R^2$ of 0.9994 and a root mean square error (RMSE) of 0.0466 $W\,m^{-1}\,K^{-1}$. The ETR model also generalized well to twelve previously unseen (randomly chosen) low and high $\kappa_L$  compounds with diverse space group symmetries, reaching an $R^2$ of 0.961 against DFT benchmarks. Notably, the model excels in predicting $\kappa_L$ for both low- and high-symmetry compounds, enabling efficient high-throughput screening. We also demonstrate this capability by screening ultralow and ultrahigh $\kappa_L$ candidates among 960 half-Heusler \textcolor{black}{compounds} and 60,000 ICSD compounds from the AFLOW database. This result shows reliability of model developed for screening of potential thermoelectric materials. At the end, we have tested model's prediction ability on systems that have experimental $\kappa_L$ available that shows model's ability to search material that has desirable experimental $\kappa_L$ for thermoelectric applications.

   
\end{abstract}

\maketitle

\section{Introduction}

\textcolor{black}{With the growing demand for green energy solutions, there has been a significant surge in the discovery of new materials with tailored thermal transport properties. One of the most abundant energy sources is heat, and its control and conversion particularly through thermoelectric technology depend critically on lattice thermal conductivity ($\kappa_L$).} A key material property governing the efficiency of thermoelectric devices is the lattice thermal conductivity ($\kappa_L$), which directly affects heat transport. This property is critical for various applications, including thermoelectrics for power generation \citep{zheng2008thermoelectric}, thermal transducers \citep{morelli2006high}, heat management in electronic devices, and energy storage systems \citep{CAI2019238}.
Identifying materials with either extremely high or low $\kappa_L$ is a computationally intensive and time-consuming task, primarily due to the need for phonon-based calculations. In any {\it ab-initio} simulation, these calculations are among the most resource-demanding steps and are highly sensitive to the symmetry and complexity of the material's crystal structure\citep{PhysRevE.108.045302,phonon_calc}. Accurate prediction of $\kappa_L$ is essential for the rational design of materials with tailored thermal properties.
Traditionally, computational approaches such as density functional theory (DFT) and molecular dynamics (MD) simulations \citep{MD} have been used to estimate $\kappa_L$. While these methods offer high accuracy, they come with significant computational costs. Within the DFT framework, the most widely used approach to calculate $\kappa_L$ involves solving the Boltzmann transport equations (BTE) \citep{BTE_AJain}. The computational demand of these simulations escalates with decreasing crystal symmetry, due to the corresponding increase in the number of atoms in the unit cell.
Although DFT-based methods provide a detailed, all-electron description capable of capturing a wide range of physical phenomena, their high computational cost limits their scalability. This bottleneck poses a major challenge in high-throughput screening of large materials databases for novel thermoelectric candidates. As a result, more efficient approaches are needed to enable large-scale exploration of materials with desirable thermal transport properties.


In recent years, machine learning (ML) has emerged as a powerful tool to accelerate materials discovery and property prediction \citep{D0NA00388C,karande_strategic_2022,hu_realistic_2024}. ML models, trained on large databases of computational or experimental data, have demonstrated high accuracy in predicting various atomic and electronic properties of materials, such as band gap\citep{li_md-hit_2024,goodall_predicting_2020,hu_realistic_2024}, formation energy\citep{li_md-hit_2024,allen_machine_2022}, magnetic moment\citep{li_md-hit_2024}, total energy\citep{li_md-hit_2024}, charge density\citep{li_md-hit_2024}, and potential energy\citep{zhong_explainable_2022}. Beyond electronic properties, ML approaches have also been employed to predict mechanical\citep{li_md-hit_2024,huang_application_2023}, thermal\citep{li_md-hit_2024,zhong_explainable_2022}, and optical\citep{huang_application_2023} properties of materials.

However, studies specifically targeting the prediction of temperature-dependent lattice thermal conductivity ($\kappa_L$) using ML remain relatively limited. Early efforts include the work of Juneja {\it et al.}, who employed Gaussian process regression to model $\kappa_L$ at room temperature (300 K)\citep{Juneja2019}. Similarly, Wang {\it et al.} used an XGBoost model to predict room-temperature $\kappa_L$ values across a dataset of materials\citep{wang_identification_2020}. More recently, some studies have addressed the prediction of $\kappa_L$ as a function of temperature. For instance, Jaafreh {\it et al.} developed a random forest-based model capable of predicting $\kappa_L$ at arbitrary temperatures\citep{jaafreh_lattice_2021}, while Li {\it et al.} proposed a neural network-based framework to model temperature-dependent $\kappa_L$ with improved flexibility and accuracy\citep{li_machine_2025}.
However, in all these studies, the training datasets are typically confined to specific classes of materials. Such limited and domain-specific datasets often result in models that lack generalization capability and exhibit bias toward the material class on which they were trained. Consequently, these models struggle to make accurate predictions across chemically or structurally diverse compounds. This highlights the pressing need to develop more generalized ML models trained on diverse and representative datasets encompassing a broad range of materials. 
Such models would possess improved transferability and robustness, enabling reliable prediction of $\kappa_L$ across different material families. 


\begin{figure*}[t]
     \centering
      \includegraphics[width=0.95\linewidth]{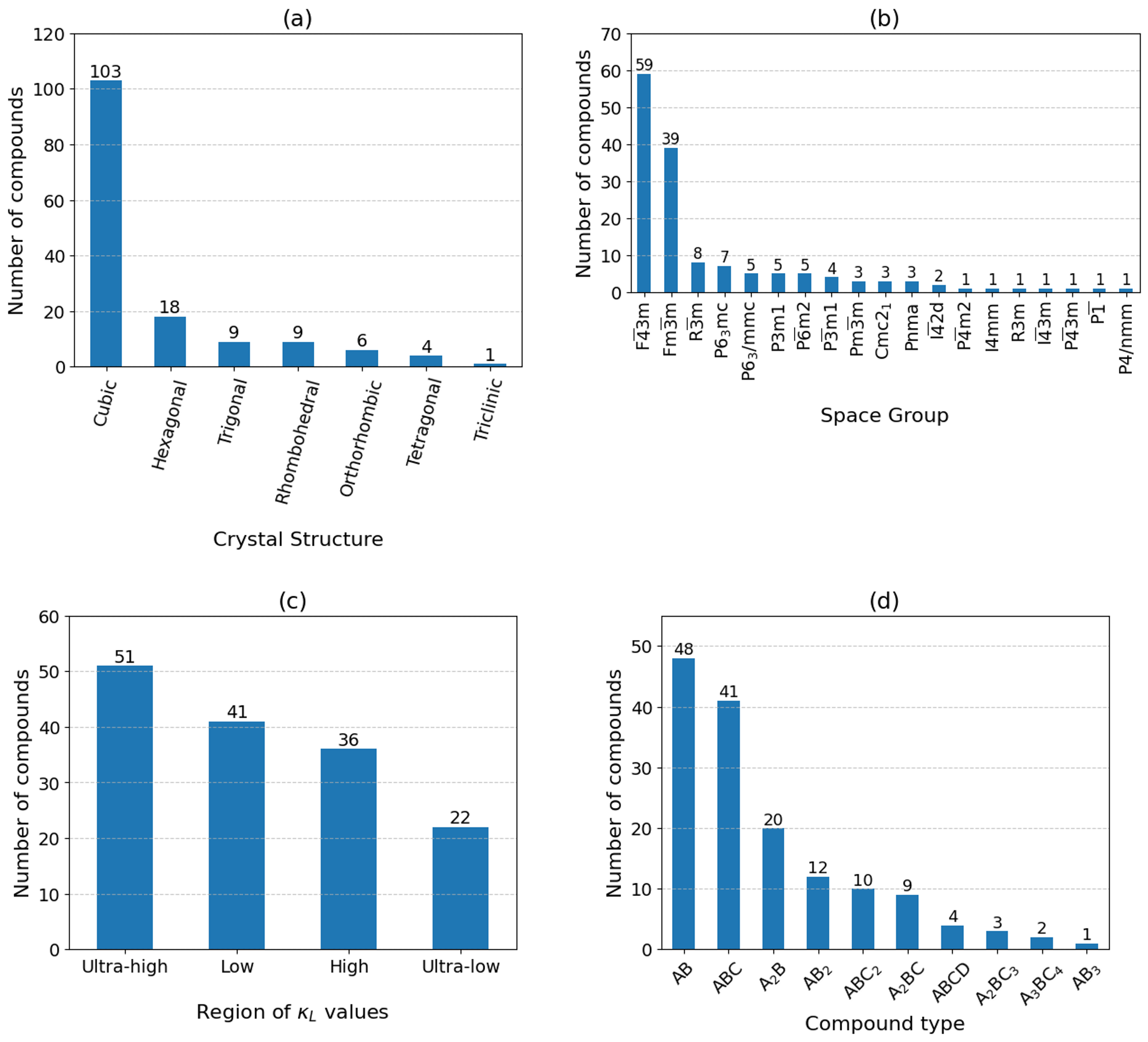}
      \caption{\small\justifying\protect\label{fig1}The distribution within the dataset of 150 compounds categorized with respect to  (a) Crystal structure \-: highlighting the prevalence of different crystal symmetry such as cubic, orthorhombic, and tetragonal (b) Space group \-: reflecting the diversity in crystallographic symmetry (c) Lattice thermal conductivity ($\kappa_L$) type \- classified into four categories based on $\kappa_L$ values at 300 K (Ultra-high: $\kappa_L > 15$ $W\,m^{-1}\,K^{-1}$; High: $5 < \kappa_L \leq 15$ $W\,m^{-1}\,K^{-1}$; Low: $1 < \kappa_L \leq 5$ $W\,m^{-1}\,K^{-1}$; Ultra-low: $0 < \kappa_L \leq 1$ $W\,m^{-1}\,K^{-1}$ (d) Compounds type \-: showing the frequency of various compositional types (e.g., AB, A$_3$BC$_4$, etc.) }
\end{figure*}

In this work, we investigate a range of nonlinear regression-based ensemble ML models to predict the temperature-dependent lattice thermal conductivity ($\kappa_L$) of a variety of materials. The models are trained on a diverse dataset comprising variable temperature $\kappa_L$ of 150 compounds. For materials representation, we utilize the MAGPIE Java library \citep{Ward2016}, which converts the compositional and structural information into various crystal feature vectors of length 271. This crystal feature vector is generated by using the elemental properties of constituent atoms (e.g., atomic weight, atomic number, melting point, mendeleev number etc.) and their positions in the unit cell (a full list of various compositional and structural features are listed in Sec. II of the supporting information (SI)\cite{supplementary_file}).  Each compound, characterized by its unique chemical composition and space group, corresponds to a distinct feature vector that is used for training and validation of the ML models.
To improve model performance and reduce dimensionality, we employ a feature selection methods\citep{mark_a_hall_feature_nodate} to identify the most informative 53 features out of the original 272 (including Temperature) (the list of 53 features are shown in Table S6 of SI\cite{supplementary_file}). Seven different ensemble learning based regression models like Decision Tree\citep{breiman1984classification}, Random Forest\citep{breiman2001randomforest}, XGBoost\citep{chen2016xgboost}, Gradient Boosting Regressor\citep{friedman2001gradientboosting}, Decision Tree Adaboost\citep{freund1997boosting}, Decision Tree Bagging \citep{breiman1996bagging}, and Extra Trees Regressor\citep{geurts2006extratrees} are trained and validated. These trained models are evaluated using three performance metrics: the coefficient of determination ($R^2$), mean absolute error (MAE), and root mean square error (RMSE). Among these models, the Extra Trees Regressor (ETR) achieves the highest performance with an $R^2$ score of 0.9994, and average test MAE/RMSE of 0.0249/0.0466 $W\,m^{-1}\,K^{-1}$, establishing it as the best-performing model. Additionally, we have analyzed the feature importance scores across all models to understand the relative contribution of individual features toward accurate $\kappa_L$ prediction. The ETR model is further validated by predicting $\kappa_L$ values for randomly selected compounds reported in the literature and comparing them with corresponding DFT-computed results. Finally, we demonstrate the utility of our approach in high-throughput screening by applying the ETR model to a dataset of 960 half-Heusler \textcolor{black}{compounds}, identifying both low and high $\kappa_L$ candidates. We further extend the screening to ~60,000 inorganic compounds from the ICSD database accessed via the AFLOW library\citep{aflow}, identifying promising ultra-low and ultra-high $\kappa_L$ candidate materials useful for specific applications. At the end, we have compared the ML model predictions with the experimental $\kappa_L$ values for understanding the reliability of model to search ultralow $\kappa_L$ compounds. 
\textcolor{black}{In contrast to previous studies that were restricted to narrow material classes, our work bridges this gap by developing a generalized, physics-informed ML framework trained on a chemically and structurally diverse dataset, thereby enabling accurate prediction of $\kappa_L$ across a broad materials space.}


\section{Methods}
\subsection{Dataset preparation}


To construct a reliable and diverse dataset for machine learning, we compiled DFT-computed lattice thermal conductivity ($\kappa_L$) values for 150 crystalline compounds across a range of temperatures (100 \-- 1000 K) from various literature sources \textcolor{black}{\citep{Juneja2019,FanMo2024,GOVINDARAJ2023,HE2022,zhang2024,SHARMA2023106289,wang2019,Andrea_2015,wei2023,RUGUT2021,YANG2023,Gandi2017,LUO2023,SINGH2017120,JASPAL2023161,PARVIN2021104068,ANAND20191226,zhu_significant_2020,YOU2023101015,CuBiTeO,haque2021}}, resulting in a total of \textcolor{black}{4127} data entries for training. All $\kappa_L$ values were obtained using first-principles calculations based on the linearized phonon Boltzmann transport equation (BTE) solved through an iterative approach, ensuring high fidelity in the dataset.
To assess the dataset\textquotesingle s diversity, we analyzed the 150 compounds in terms of space group symmetry, chemical formula, constituent elements, crystal structure, and the range of $\kappa_L$ values. The dataset encompasses binary, ternary, and quaternary compounds, ensuring coverage across a broad chemical space (Refer to Table S1 and Table S2 in the SI file for \textcolor{black}{150} compounds list). For a comprehensive visualization, the element-wise distribution of compounds is presented in Fig. S1(a), where the occurrence frequency of each element is represented on a periodic table using a color bar \citep{periodictableplot}. Additionally, a histogram in Fig. 1S(b) illustrates the element distribution more explicitly. The analysis reveals that elements such as Se, Te, S, Sb, and Sn are highly represented in the dataset, while elements like Ce, H, La, Zn, Lu, Ru, F, Sm, O, Zr, and Ta appear less frequently.

Upon further analysis, the dataset is found to comprise compounds spanning ten distinct compositional types. Among these, 48 compounds exhibit an AB-type composition, while 2 compounds follow an A$_3$BC$_4$-type composition, as shown in Fig.~\ref{fig1}(d). The crystal structure distribution (Fig.~\ref{fig1}(a)) reveals that the majority of compounds belong to the cubic structures (with space groups $F\overline{4}3m$ and $Fm\overline{3}m$), with only a few belonging to orthorhombic/tetragonal/triclinic structures. A wide range of space groups is observed across the dataset, indicating the presence of both high and low-symmetry structures (Fig.~\ref{fig1}(b)). Additionally, the compounds are categorized into four groups i.e. Ultra-high, High, Low, and Ultra-low based on the values of their lattice thermal conductivity ($\kappa_L$) at 300 K, as described in Fig.~\ref{fig1}(c). Overall, Fig.~\ref{fig1} highlights the diversity of the dataset in terms of chemical composition, symmetry, crystal structure, and lattice thermal conductivity values.

\begin{figure*}[t]
     \centering
      \includegraphics[width=1.0\linewidth]{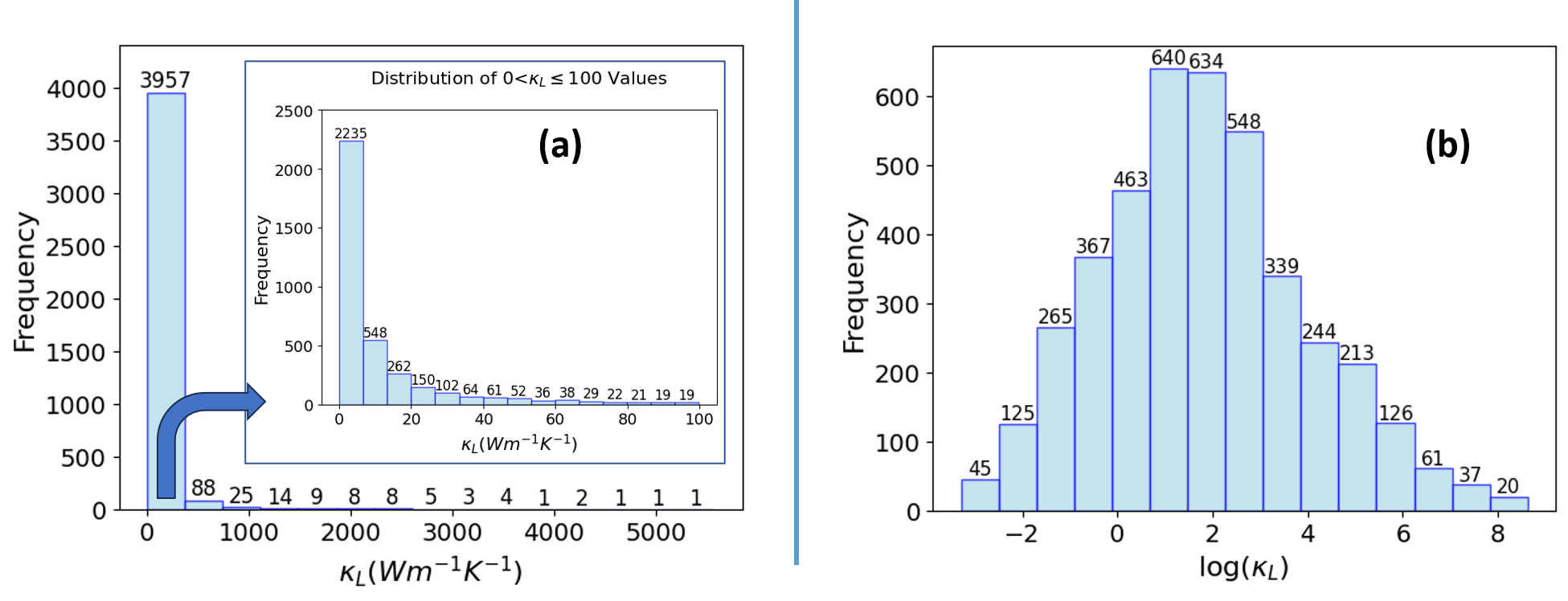}
      \caption{\small\justifying\protect\label{fig2} 
 Distribution of 4127 entries in the dataset: (a) $\kappa_L$ across all temperatures (inset: zoomed view of $\kappa_L \leq 100$ $W\,m^{-1}\,K^{-1}$), showing a highly skewed distribution; (b) log($\kappa_L$), revealing an approximately normal distribution suitable for machine learning modeling. }
\end{figure*}

The analysis of lattice thermal conductivity ($\kappa_L$) is performed to understand its distribution across all compounds in the dataset. Fig.~\ref{fig2}(a) displays the distribution of 4127 data points, showing that $\kappa_L$ spans several orders of magnitude. For instance, CsK$_2$Sb (space group: Fm$\overline{3}$m) exhibits an extremely low $\kappa_L$ of just 0.037 $W\,m^{-1}\,K^{-1}$ at 1000 K, while SiC (space group: $F\overline{4}3m$) shows a remarkably high value exceeding 5000 $W\,m^{-1}\,K^{-1}$ at 100 K. Moreover, the inset of Fig.~\ref{fig2}(a) reveals that a significant number of compounds around 2135 have their $\kappa_L$-values lie within a narrow range of 0 to 7 $W\,m^{-1}\,K^{-1}$. This highly skewed distribution can negatively impact the performance of machine learning (ML) models by introducing bias towards more frequently occurring values and hindering the model\textquotesingle s generalization ability. To address this problem, the logarithm of $\kappa_L$, i.e., log($\kappa_L$) is used instead of their raw values. This transformation converts the skewed distribution into an approximately normal distribution, as shown in Fig.~\ref{fig2}(b), which is more suitable for ML training. Using log-transformed $\kappa_L$ not only enhances model performance by reducing bias but also facilitates faster and more stable training.

\subsection{Features analysis}

\begin{figure*}[t]
     \centering
      \includegraphics[width=0.9\linewidth]{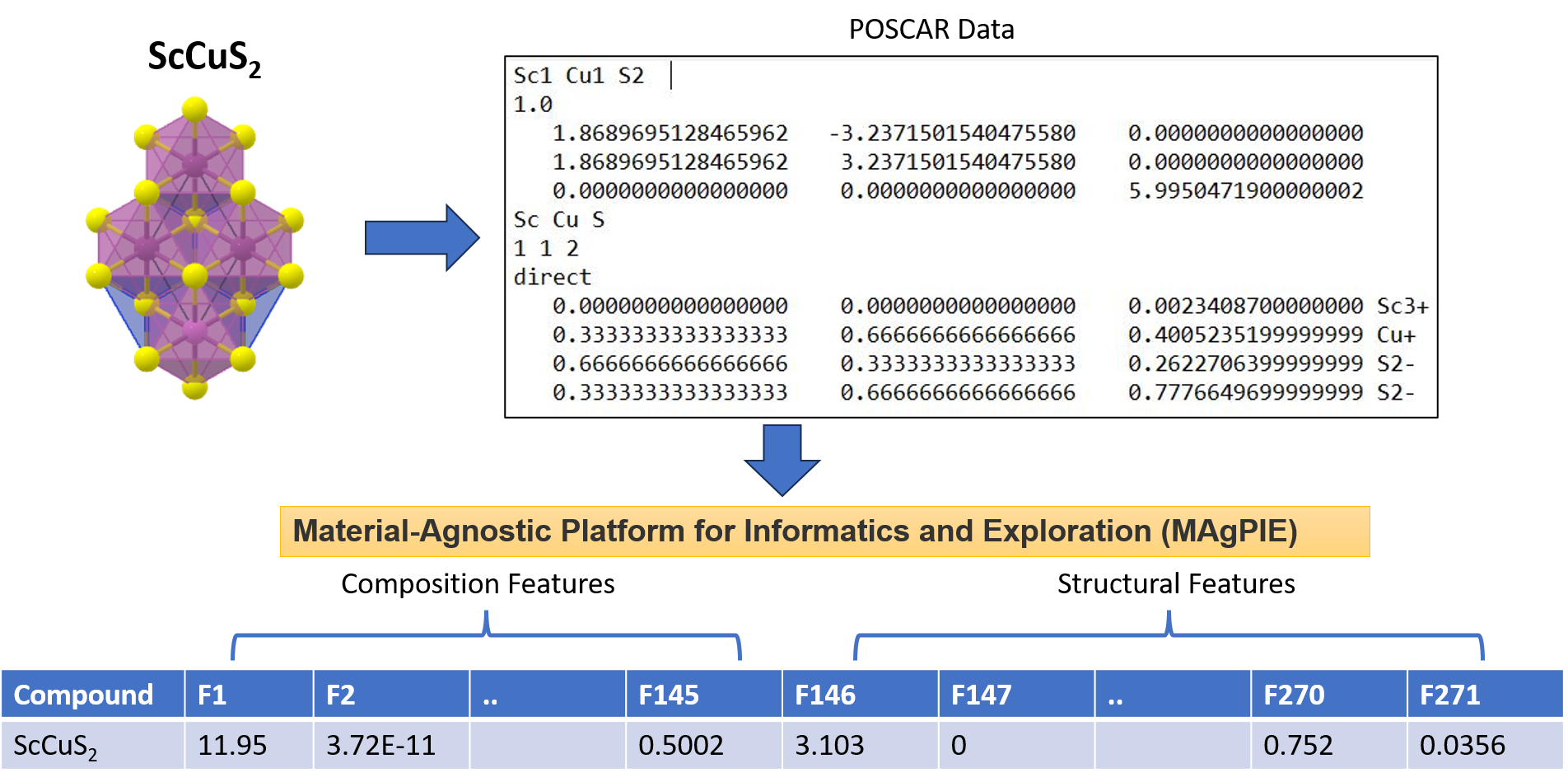}
      \caption{\small\justifying\protect\label{fig3} Feature generation work flow of a test compound, ScCuS$_2$, using MagPie Library\citep{Ward2016}. Further details of the elemental properties and the associated statistical descriptors used to generate these features is provided in Table S3 of SI\cite{supplementary_file}.}
\end{figure*}

Predictive power of any ML model is highly dependent on input features of training data. In this work, two categories of features were used: crystal features and composition features. These features were generated on the basis of crystal structure and composition of compounds that were used to build training dataset (see details in Table S3, Table S4 and Table S5 of SI\cite{supplementary_file}). In the present work, feature generation is carried out using the Materials Agnostic Platform for Informatics and Exploration (MAgPIE) \citep{Ward2016}. This platform extracts structural and compositional features directly from the structure file (e.g. POSCAR in VASP package) of each compound (further details on the feature generation process are provided in (SI)\cite{supplementary_file}). These structure files usually contain essential structural information of compounds such as lattice geometry, neighbouring atoms and atomic positions etc. As a core input for DFT calculations, it captures the crystal structure in detail, making it well-suited for establishing structure property relationships in materials. The overall workflow for feature generation from the structure file (POSCAR in the present case) is illustrated in Fig.~\ref{fig3}. Additionally, lattice thermal conductivity ($\kappa_L$) values were extracted from $\kappa_L$ vs. temperature plots reported in the literature using the WebPlotDigitizer tool\citep{webplot}.

\begin{figure*}[t]
     \centering
      \includegraphics[width=1\linewidth]{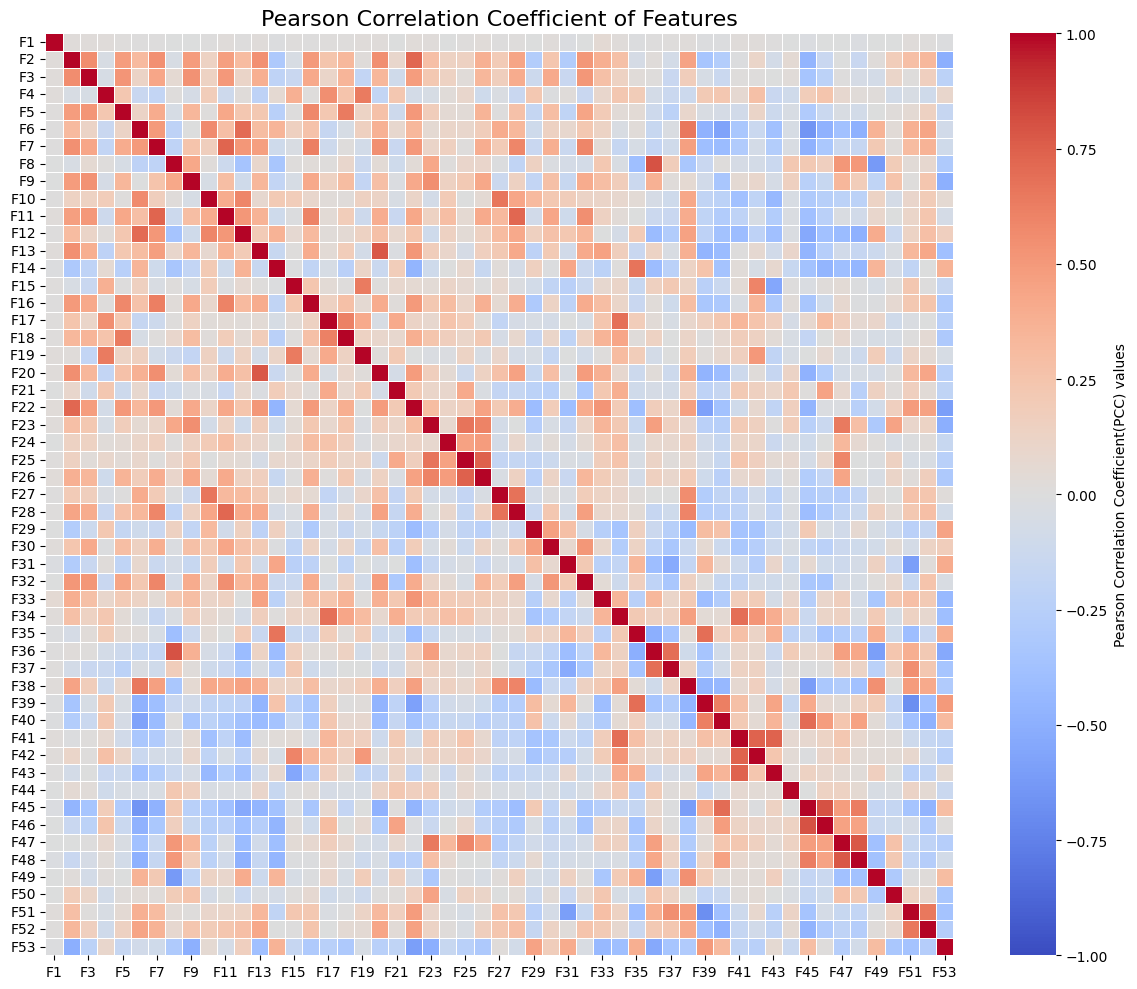}
      \caption{\small\justifying\protect\label{fig4} Pearson correlation map of 53 features used for model training. Abbreviation of all features are provided in Table-S6\cite{supplementary_file}. \textcolor{black}{The color bar represents Pearson Correlation Coefficient (PCC).}}
\end{figure*}

In Fig.~\ref{fig3}, an example compound, ScCuS$_2$, is shown to illustrate the process of extracting compositional and structural features from a POSCAR file. The resulting feature vector has a length of 271 and incorporates both elemental and structural descriptors. Each compound with a valid POSCAR file is transformed into such a feature vector, capturing its unique crystal structure and chemical composition. These features are derived from elemental properties of the constituent elements and the space group information. A complete list of the elemental properties and the associated statistical descriptors used to generate these features is provided in the Table no. S3, S4 and S5 of SI\cite{supplementary_file}.
To incorporate temperature dependence into our models, the temperature (in Kelvin) at which each $\kappa_L$ value is evaluated is included as an additional feature, bringing the total input feature vector length to 272. Including temperature as a feature allows the model to learn how $\kappa_L$ evolves across a wide range of temperatures (typically above 100 K), enabling more accurate and physically consistent predictions. This also facilitates direct comparison with DFT-calculated $\kappa_L$ values reported in the literature. Moreover, combining temperature with structural and compositional features allows the model to better capture the interactions and dependencies that govern thermal transport in materials, as supported by prior work on feature interactions in machine learning models \citep{friedman_predictive_2008}.
To reduce model complexity and improve generalizability, feature selection was performed using a combination of standard filtering techniques: variance thresholding and Pearson correlation analysis\citep{kamalov_mathematical_2025}. In the first step, features exhibiting very low variance across samples (variance < 0.16) were considered nearly constant and therefore uninformative; 64 such features were removed. The remaining features were then screened using Pearson correlation, and highly correlated features (correlation coefficient > 0.80) were excluded to avoid redundancy. After both filtering stages, 53 informative and non-redundant features were retained for subsequent model training.
These selected features are visualized in the Pearson correlation heatmap shown in Fig.~\ref{fig4}, with corresponding abbreviations defined in Table S6\cite{supplementary_file}. A comprehensive summary of the full feature set and the filtering process is also provided in Table S6 and Fig. S1 of SI\cite{supplementary_file}. These statistical features were later used to train and evaluate several ensemble learning regression models, including Decision Tree, Random Forest, Gradient Boosting, eXtreme Gradient Boosting, AdaBoost with Decision Tree, Bagging Regressor, and Extra Trees Regressor.

\textcolor{black}{It is important to note that all $\kappa_L$(T) data points across the range of 100{\em --}1000 K were pooled together during training, rather than training separate models at each temperature. To enable temperature-resolved predictions, the temperature (in Kelvin) corresponding to each $\kappa_L$ entry was explicitly incorporated as an additional input feature in the descriptor set. This strategy allows the model to simultaneously learn both the compositional/structural dependence and the temperature dependence of $\kappa_L$. As a result, for any new compound, the trained model can be queried at a specific temperature (e.g., 300 K) or across the full temperature window (100{\em --}1000 K) to yield $\kappa_L$ predictions consistent with the trends learned during training.}

\setlength{\tabcolsep}{10pt}
\begin{table*}[t]
\caption{\small\justifying\protect\label{tab:tabel1} Comparison of evaluation metrics for predicting log scale lattice thermal conductivity ($\kappa_L$) using seven machine learning models after repeated K-fold cross-validation. The performance is assessed using the evaluation metrics calculated on the log-transformed $\kappa_L$ values.}
\centering{}%
\begin{tabular}{ccccc}
\toprule 
\multirow{2}{*}{ML model} & \multirow{2}{*}{$R^2$ (test set)      } & \multirow{2}{*}{RMSE (test set)     } & \multirow{2}{*}{MAE (test set)       } & \multirow{2}{*}{Time cost}\tabularnewline
 &      &      &      & \tabularnewline
\midrule
\midrule 
Extra Trees  & 0.9994 & 0.0466 & 0.0249 & 3.33 min\tabularnewline
Random Forest  & 0.9982 & 0.0850 & 0.0484 & 5.05 min\tabularnewline
DT-AdaBoost  & 0.9981 & 0.0898 & 0.0724 & 3.12 min\tabularnewline
XGBoost  & 0.9980 & 0.0928 & 0.0632 & 82.58 min\tabularnewline
DT-Bagging & 0.9976 & 0.1003 & 0.0593 & 0.54 min\tabularnewline
Decision Tree (DT) & 0.9957 & 0.1318 & 0.0852 & 0.08 min\tabularnewline
Gradient Boosting  & 0.9844 & 0.2636 & 0.2046 & 1.64 min\tabularnewline
\bottomrule
\end{tabular}
\end{table*} 

\subsection{Machine Learning Models}
In this study, we employed seven supervised ensemble-based regression algorithms {\em --} Decision Tree\citep{breiman1984classification}, Random Forest\citep{breiman2001randomforest}, Gradient Boosting\citep{friedman2001gradientboosting}, Extreme Gradient Boosting (XGBoost)\citep{chen2016xgboost}, AdaBoost with Decision Trees\citep{freund1997boosting}, Bagging Regressor\citep{breiman1996bagging}, and Extra Trees Regressor\citep{geurts2006extratrees} all implemented using the scikit-learn\citep{sklearn} python library and trained using default hyperparameters of respective models. These models were selected because of their robustness in handling nonlinear relationships and their proven success in materials informatics. Details of algorithmic principles can be found in the cited references and SI file. \textcolor{black}{For model training and testing purpose, dataset is divided into 80\% training and 20\% testing data. Distribution of train-test data of log($\kappa_L$) is provided in Fig. S5\citep{supplementary_file}.}

After training, the performance of each model was evaluated to determine the most suitable one for predicting $\kappa_L$. This evaluation was based on widely used regression metrics, including Root Mean Square Error (RMSE), Coefficient of Determination ($R^2$), and Mean Absolute Error (MAE). These metrics are defined as follows:

(a) Root Mean Square Error (RMSE): 

\begin{equation}
\text{RMSE} = \sqrt{\frac{1}{n} \sum_{i=1}^{n} (y_i - \hat{y}_i)^2}\label{eq:rmse}
\end{equation}

(b) Coefficient of determination ($R^2$): 

\begin{equation}
R^2 = 1 - \frac{\sum_{i=1}^{n} (y_i - \hat{y}_i)^2}{\sum_{i=1}^{n} (y_i - \bar{y})^2}\label{eq:r2}
\end{equation}

(c) Mean Absolute Error (MAE):
\begin{equation}
\text{MAE} = \frac{1}{n} \sum_{i=1}^{n} \left| y_i - \hat{y}_i \right|\label{eq:mae}
\end{equation}

where
\begin{itemize}
    \item \( y_i \) - actual values of the dependent variable.
    \item \( \hat{y}_i \) - predicted values from the regression model.
    \item \( \bar{y} \) - mean of the actual values.
    \item \( n \) - total number of data points.
\end{itemize}

where $i$ specify the $i^{th}$ material sample in the dataset. In addition to above metrics, running time of the models is also taken into account to measure the computational cost of different ML models. 

The performance of the models was assessed using the repeated K-fold cross-validation technique, a robust model evaluation approach widely adopted in supervised learning tasks. In standard K-fold cross-validation, the dataset is partitioned into K equal subsets (folds). The model is trained on (K\- - 1) folds and validated on the remaining fold. This process is repeated K times, with each fold serving as the validation set once.
To further enhance the reliability of performance estimates, we employed repeated K-fold cross-validation, in which the entire K-fold process is repeated multiple times with different random splits of the data. This approach reduces variance in the evaluation metrics and provides a more comprehensive understanding of the model\textquotesingle s generalization capability.
The use of repeated K-fold cross-validation in this work ensures that the model performance metrics reflect a stable average across different data partitions, minimizing the effects of randomness and overfitting. This method was applied consistently across all regression models during training and testing phases.
The comparative performance metrics obtained through this evaluation including RMSE, $R^2$, MAE, and training time are summarized and analyzed in Table \ref{tab:tabel1}.

\textcolor{black}{In each cross-validation cycle, approximately 80\% of the dataset ($\simeq$ 3288 points) was used for training and 20\% ($\simeq$ 822 points) for testing. This process was repeated ten times with random shuffling to ensure that all 4127 data points were included in both training and validation. The average training errors for the Extra Trees Regressor were $R^2=$0.9996, RMSE= 0.041 $W\,m^{-1}\,K^{-1}$, and MAE=0.021 $W\,m^{-1}\,K^{-1}$, while the corresponding test errors were $R^2=$ 0.9994, RMSE= 0.0466 $W\,m^{-1}\,K^{-1}$, and MAE = 0.0249 $W\,m^{-1}\,K^{-1}$.
}

\section{Results and Discussions}
\subsection{Model Training and Evaluation}
Modeling the temperature-dependent behavior of $\kappa_L$ was performed using a suite of supervised machine learning (ML) algorithms, including Decision Tree\citep{breiman1984classification}, Random Forest\citep{breiman2001randomforest}, Gradient Boosting\citep{friedman2001gradientboosting}, Extreme Gradient Boosting (XGBoost)\citep{chen2016xgboost}, AdaBoost with Decision Trees\citep{freund1997boosting}, Bagging Regressor with Decision Trees\citep{breiman1996bagging}, and Extra Trees Regressor\citep{geurts2006extratrees}. These methods are ensemble learning techniques, known for improving prediction accuracy by combining multiple base learners. Such approaches are increasingly employed in materials informatics due to their robustness and adaptability to complex datasets\citep{butler_machine_2018}.
To train the ML models, a dataset comprising the log($\kappa_L$) values (target property) and the preprocessed structural and compositional features was used. The training and evaluation process involved 10-fold cross-validation i.e. repeated 10 times with random shuffling of the dataset in each repetition. This strategy ensures a robust and unbiased estimate of model performance by minimizing variance and avoiding overfitting.
Model evaluation was based on three widely used regression performance metrics: the coefficient of determination ($R^2$), mean absolute error (MAE), and root mean square error (RMSE). These metrics were computed for each fold and repetition, and the average values across all iterations were used for comparative analysis.

\begin{figure*}
\begin{centering}
\includegraphics[scale=0.77]{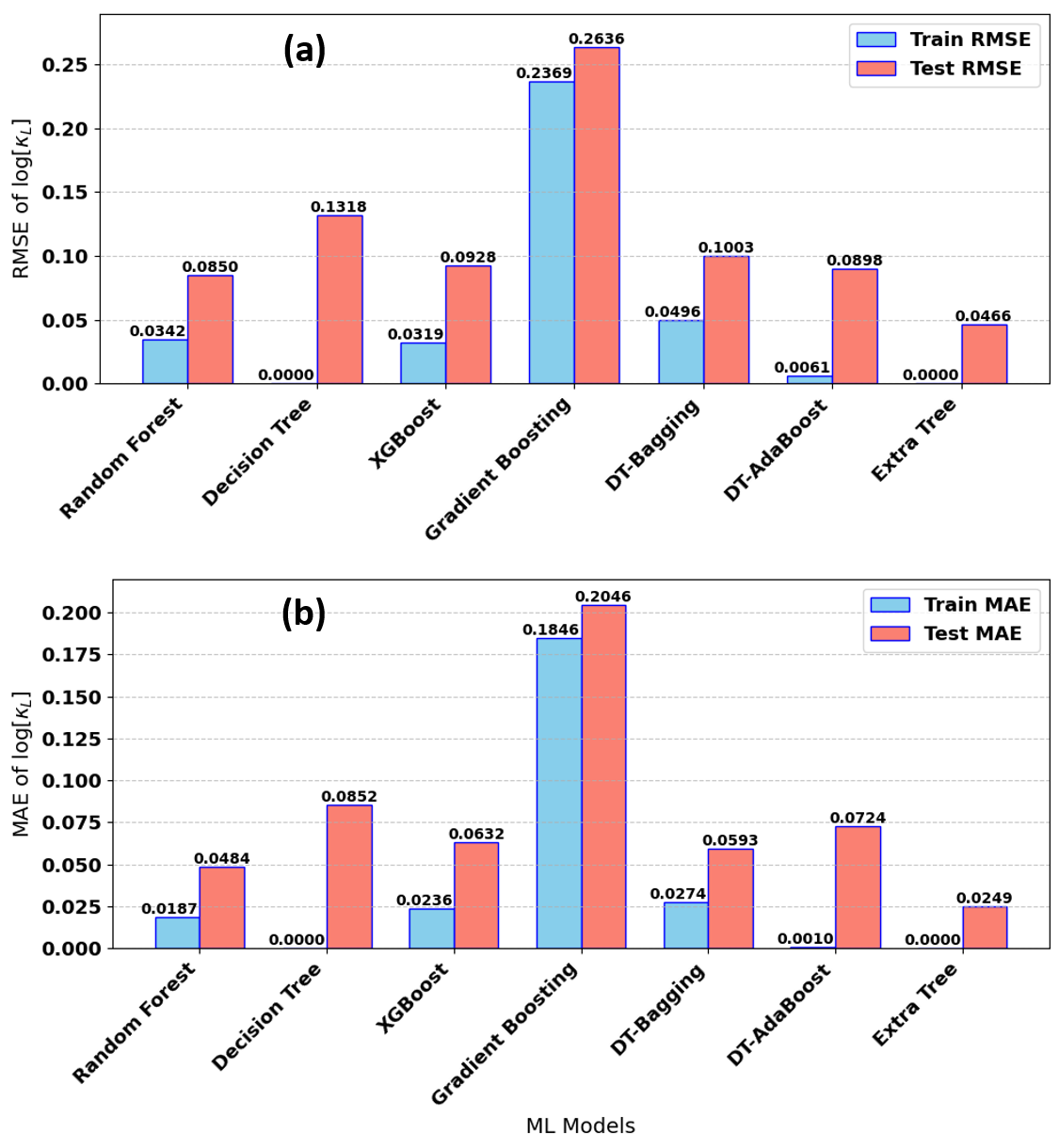}
\par\end{centering}
\caption{\small\justifying\protect\label{fig5} Comparison of (a) root mean square error (RMSE) and (b) mean absolute error (MAE)  for different ML models used in predicting the logarithm of $\kappa_L$. }
\end{figure*}

Table \ref{tab:tabel1} summarizes the average performance metrics of various machine learning (ML) models, evaluated using repeated K-fold cross-validation. This cross-validation strategy provides a statistically robust framework for comparing the generalization capabilities of different models across multiple train-test splits, offering deeper insights into the model performance\citep{raschka_model_2020}. The comparative performance plot in Fig.~\ref{fig5} illustrates the MAE and RMSE values for both training and testing phases across all ensemble-based ML models employed in this study. These include models such as Decision Tree, Random Forest, Gradient Boosting, XGBoost, Decision Tree AdaBoost, Decision Tree Bagging, and Extra Trees Regressor. Based on this rigorous evaluation, the Extra Trees Regressor demonstrated superior predictive performance, achieving the highest average $R^2$ score of 0.9994 and the lowest average RMSE and MAE scores of 0.0479 and 0.0249 $W\,m^{-1}\,K^{-1}$, respectively, for the log-transformed $\kappa_L$ values. In contrast, the Gradient Boosting Regressor yielded the lowest $R^2$ score of 0.9841 and the highest RMSE/MAE values of 0.2644/0.2054 $W\,m^{-1}\,K^{-1}$, indicating comparatively weaker generalization performance. 
The time cost reported in Table \ref{tab:tabel1} corresponds to the total duration required to perform 10-fold cross-validation on a DGX-3 server (our in-house high-performance computing facility). It is important to note that these timings are highly dependent on the system and its configuration. A qualitative analysis of the time taken indicates that XGBoost is the most computationally expensive model, whereas the decision tree is the least expensive among the models considered.

\begin{figure*}[htbp]
\begin{centering}
\includegraphics[scale=0.68]{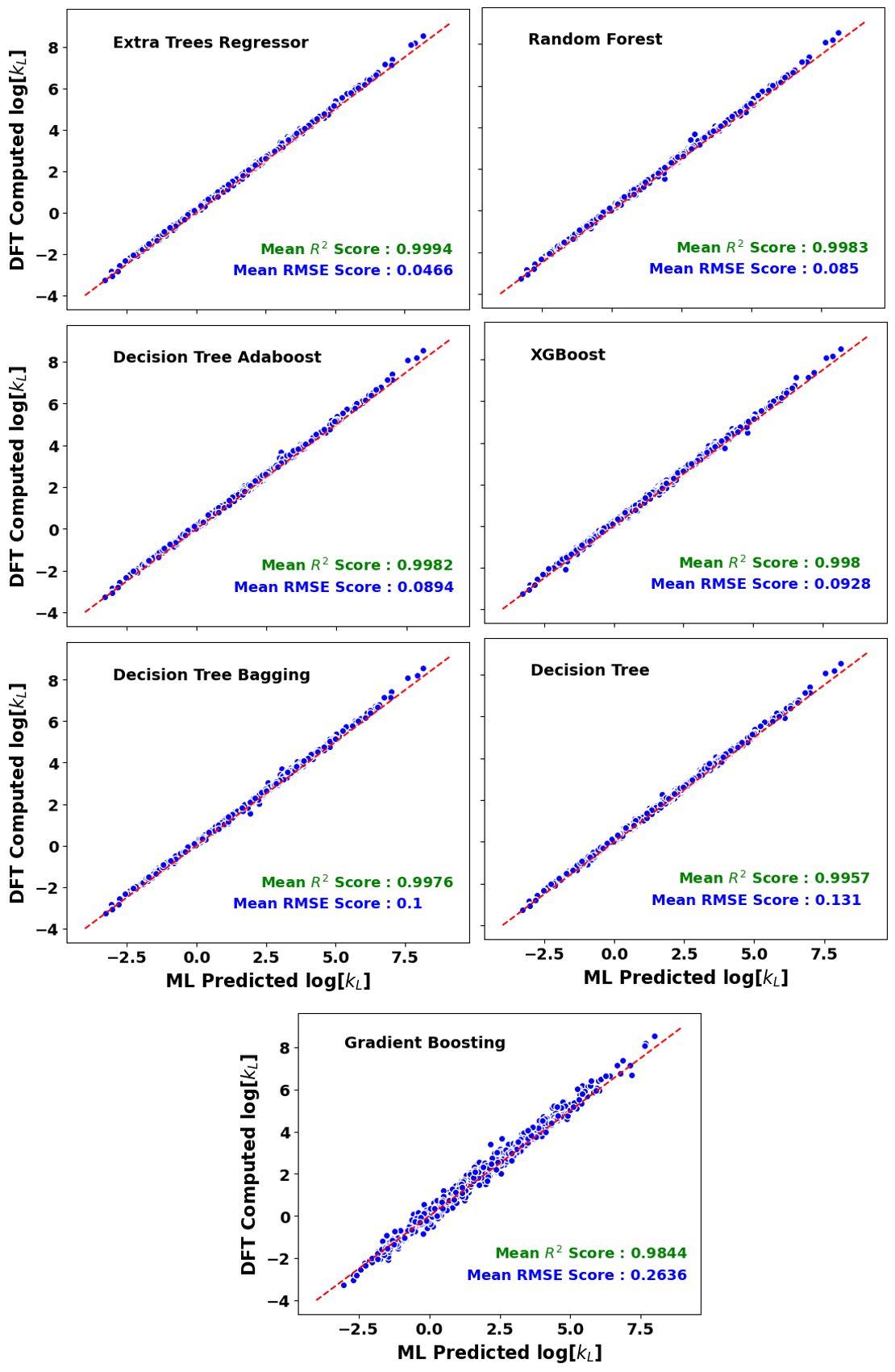}
\par\end{centering}
\caption{\small\justifying\protect\label{pair_plot_7_models} Pair plots showing the performance of different ML models on the test data. The plots compare the predicted $\kappa_L$ values obtained from ML models with those from DFT calculations. }
\end{figure*}

\begin{figure*}
\begin{centering}
\includegraphics[scale=0.37]{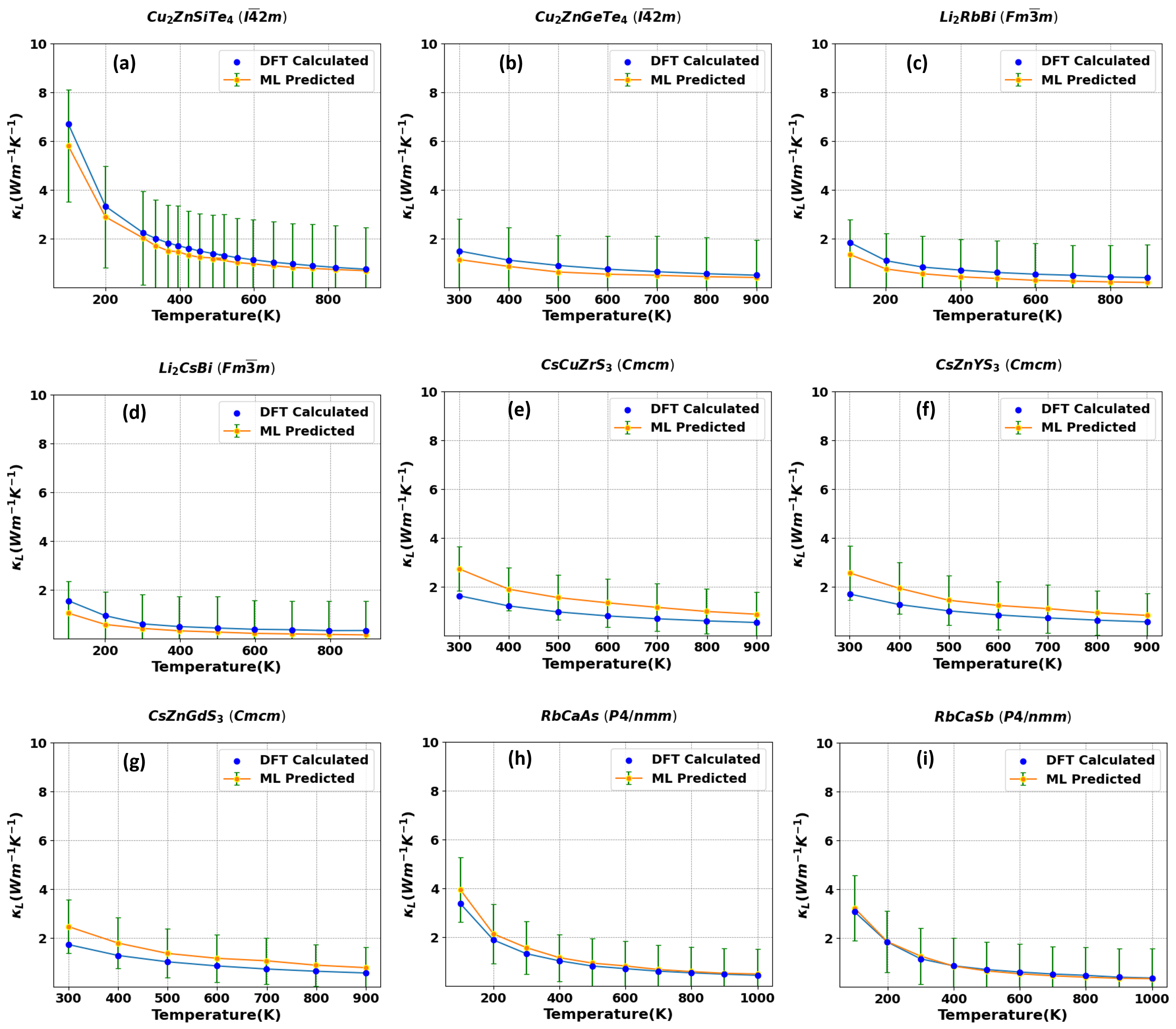}
\par\end{centering}
\caption{\small\justifying\protect\label{fig7} \textcolor{black}{
Comparison between the Extra Trees Regressor–predicted lattice thermal conductivity ($\kappa_L$) values (with error bars) and DFT results for nine low-$\kappa_L$ compounds randomly selected from the literature. Predictions were made with temperature included as an explicit input feature, allowing the model to estimate $\kappa_L$
 at any temperature within the 100–1000 K range. The green error bars represent the standard deviation of predictions from individual trees within the ensemble model.}}
\end{figure*}

These results establish the Extra Trees Regressor as the most suitable model for accurate prediction of $\kappa_L$ and for subsequent compound screening tasks. To further interpret model behavior, the top 15 most influential features identified by the Extra Trees model are shown in Fig.~\ref{pair_plot}(b). A comprehensive list of feature importance scores for all 53 features used in the final model is provided in Table S7, offering valuable insight into the structural and compositional factors that govern the prediction of $\kappa_L$. 

Fig.~\ref{pair_plot_7_models} presents the pair plots of 7 ML models, illustrating the relationship between the predicted and DFT-calculated values of lattice thermal conductivity ($\kappa_L$) on the test dataset. The plots reveal that, for most of the models, the predicted $\kappa_L$ values are in good agreement with the DFT-calculated results. Ideally, an $R^2$ value of 1 indicates perfect predictive performance i.e., the model predicted $\kappa_L$ exactly match the actual $\kappa_L$ values computed via DFT. While such perfection is rarely achieved in practical scenarios, $R^2$ values approaching 1 suggest strong predictive capability. Among the models assessed, the Extra Trees Regressor achieved the highest $R^2$ score, indicating superior accuracy, whereas the Gradient Boosting Regressor yielded the lowest $R^2$ score among the models evaluated.

It is interesting to note that almost all the machine learning models used in this study yielded excellent performance, achieving high $R^2$ scores (>0.99) along with promising RMSE and MSE values (<0.27). These results are quantitatively superior to most of the previously reported studies on ML-based prediction of $\kappa_L$\cite{wang_identification_2020,li_machine_2025,jaafreh_lattice_2021,lin_machine_2024}.  \textcolor{black}{The excellent predictive performance of the Extra Trees model (Fig.~\ref{fig5}) can be attributed to few intrinsic details of the model itself i.e. (1) Higher randomness (better generalization) (2) Lower variance without bias increase (3) Efficiency and robustness for high-dimensional descriptors. Each of these features are described in details below.  }

\begin{figure*}
\begin{centering}
\includegraphics[scale=0.39]{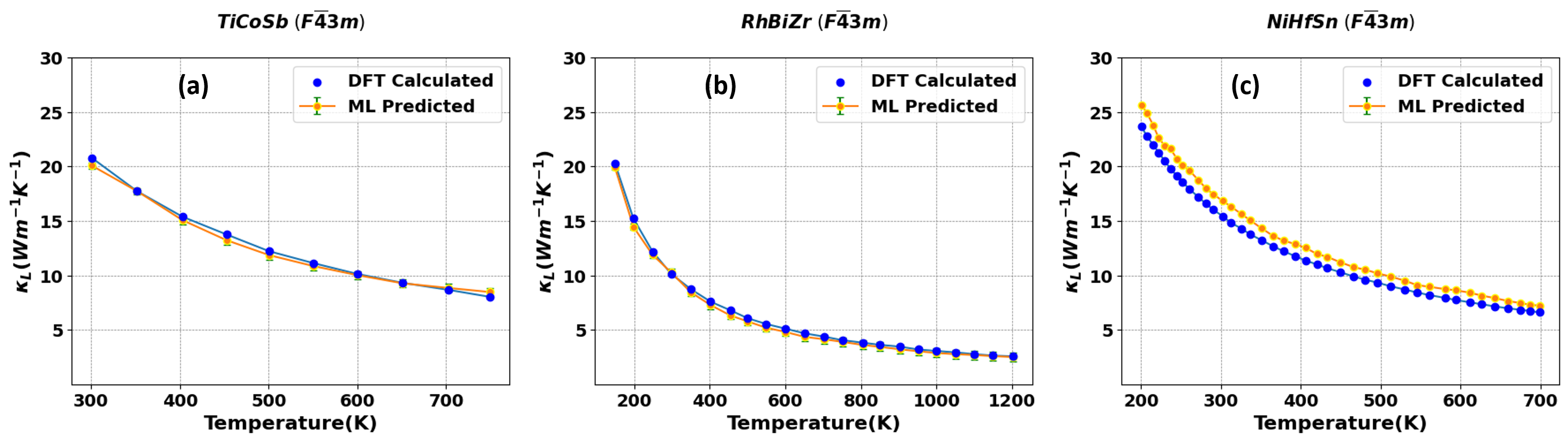}
\par\end{centering}
\caption{\small\justifying\protect\label{fig7b} \textcolor{black}{Comparison of predicted ($\kappa_L$) values from the Extra Trees Regressor with DFT results for three high-$\kappa_L$ compounds.}}
\end{figure*}


\begin{itemize}
    \item \textcolor{black}{{Higher randomness $\rightarrow$ better generalization:} 
    Unlike Random Forest (RF), which uses bootstrap sampling and optimized split thresholds, ETR selects split points completely at random for each feature. This increases model diversity and reduces correlation between trees, leading to improved generalization particularly beneficial when dealing with noisy, heterogeneous datasets such as ICSD materials data.}

    \item \textcolor{black}{{Lower variance without bias increase:} 
    The increased randomness reduces overfitting (variance) while maintaining low bias. This balance is especially advantageous for physical property prediction, where features can have correlated or nonlinear interactions that simpler models (e.g., Decision Trees, AdaBoost) often overfit.}

    \item \textcolor{black}{{Efficiency and robustness for high-dimensional descriptors:} 
    ETR can efficiently handle large, high-dimensional feature spaces without requiring heavy hyperparameter tuning (unlike XGBoost or AdaBoost). This makes it ideal for materials informatics, where descriptors derived from composition, bonding, and structure often exhibit multicollinearity.}
\end{itemize}

\begin{figure*}
\begin{centering}
\includegraphics[scale=0.43]{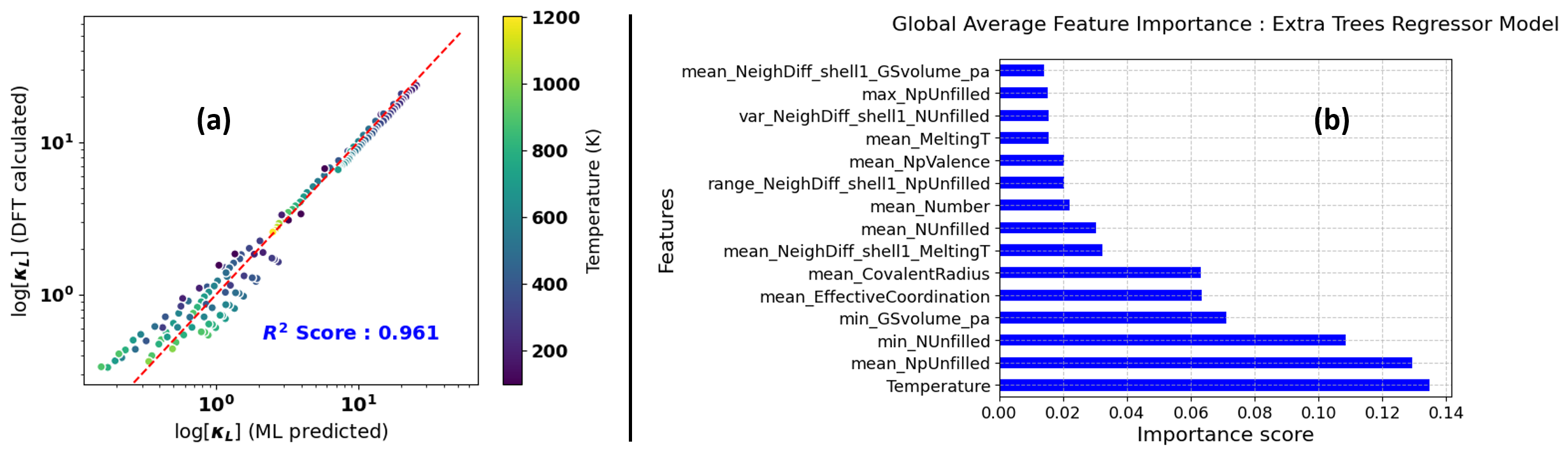}
\par\end{centering}
\caption{\small\justifying\protect\label{pair_plot}\textcolor{black}{(a) Pair plot comparing the Extra Trees Regressor–predicted and DFT-calculated lattice thermal conductivity ($\kappa_L$) values for twelve test compounds across different temperatures.
(b) Feature importance scores of the top 15 descriptors used in the Extra Trees Regressor model.}}
\end{figure*}

\subsection{Model performance on Test Compounds}
Beyond the training and test datasets, the Extra Trees Regressor model was employed to predict the lattice thermal conductivity ($\kappa_L$) of nine low $\kappa_L$ compounds and three high $\kappa_L$ compounds randomly selected from the literature, as shown in Fig.~\ref{fig7} and Fig.~\ref{fig7b} respectively. These predictions were made across a temperature range of 100-900 K for some compounds and 300-900 K for others. All selected low $\kappa_L$ compounds possess $\kappa_L$ values below 10 $W\,m^{-1}\,K^{-1}$ and high $\kappa_L$ compound posses $\kappa_L$ value below 30 $W\,m^{-1}\,K^{-1}$, enabling a focused evaluation of the model's performance in the both low $\kappa_L$ and high $\kappa_L$ regime.
For all twelve compounds, the $\kappa_L$ values predicted by the Extra Trees model exhibit good agreement with those obtained from first-principles calculations \citep{sahni_thermoelectric_2023,sharma_cu_2024,guo_comparison_2024,pal_accelerated_2021,song_thermal_2022,ANAND20191226,wei2023,Andrea_2015}. For instance, the room-temperature $\kappa_L$ of Cu$_2$ZnSiTe$_4$ (space group $I\overline{4}2m$) is calculated to be 2.24 $W\,m^{-1}\,K^{-1}$\citep{sharma_cu_2024}, which is in close agreement with the model prediction of 2.03 $W\,m^{-1}\,K^{-1}$. Across all the temperature range (200-900 K), the model\textquotesingle s predictions for Cu$_2$ZnSiTe$_4$ remain consistently close to the DFT results except a slight deviation at 100 K and 200 K.

Similarly, for Cu$_2$ZnGeTe$_4$ (space group $I\overline{4}2m$), the model predicts a room-temperature $\kappa_L$ of 1.14 $W\,m^{-1}\,K^{-1}$, compared to the DFT value of 1.50 $W\,m^{-1}\,K^{-1}$ \citep{sahni_thermoelectric_2023}. In the case of Li$_2$RbBi and Li$_2$CsBi (both in space group $Fm\overline{3}m$), the predicted $\kappa_L$ values at 300 K are 0.56 $W\,m^{-1}\,K^{-1}$ and 0.41 $W\,m^{-1}\,K^{-1}$, respectively, which are also close to the DFT values of 0.83 $W\,m^{-1}\,K^{-1}$ and 0.60 $W\,m^{-1}\,K^{-1}$\citep{guo_comparison_2024}.
For the CsCuXS$_3$ (X = Zr, Gd, Y) family of compounds (space group $Cmcm$), the model slightly overestimates the $\kappa_L$ values compared to DFT\citep{pal_accelerated_2021}, suggesting it may be capturing an upper bound in these cases. For RbCaX (X = Sb, As) compounds (space group $P4/nmm$)\citep{song_thermal_2022}, ML model predictions closely follow the DFT computed $\kappa_L$ further indicating that the model is reasonably robust in predicting $\kappa_L$ across different structural and compositional classes within the low $\kappa_L$ regime. \textcolor{black}{Interestingly, for some compounds, the relatively larger deviations are observed at low temperatures (high $\kappa_L$ region) in Fig.~\ref{fig7}. This behavior results from several factors. First, the training dataset is skewed across the 100{\em --}1000 K range (see Fig.~\ref{fig2}(a)) and low-temperature points near the lower bound are quite diverse in nature. Second, the current descriptor set (MagPie composition + structural averages, plus temperature as a scalar) does not explicitly encode several phonon transport quantities that dominate low-T behaviour of $\kappa_L$ (Debye temperature, acoustic group velocities, phonon lifetimes, and boundary/impurity scattering contributions). Third, low-T DFT/BTE $\kappa_L$ is particularly sensitive to numerical convergence (q-mesh, supercell size) and to whether extrinsic scattering channels are included. Taken together, these points explain the increased scatter at low T. However, this behaviour is not generic across all the compounds. Prediction made by ML model matches closely or slightly overestimated/underestimated in several cases. Hence, we conclude that the ML model provides a reasonably accurate $\kappa_L$ that can facilitate an idea about the range where DFT computed results most probably lie.} 

For high $\kappa_L$ compounds, TiCoSb($F\overline{4}3m$) and RhBiZr($F\overline{4}3m$), the ML predictions closely follow DFT results (refer Fig.~\ref{fig7b}(a) and (b)). In case of NiHfSn($F\overline{4}3m$), ML predictions are slightly overestimed compared to DFT computed results. In order to address any difference in the ML predictions and the DFT computed results, we also calculated the error bar for each $k_L$ calculations at each temperature. The error is simply the standard deviation at each temperature. Notice that for all the low $\kappa_L$ compounds, ML predictions lie within the error bar.

\begin{figure*}
\begin{centering}
\includegraphics[scale=0.45]{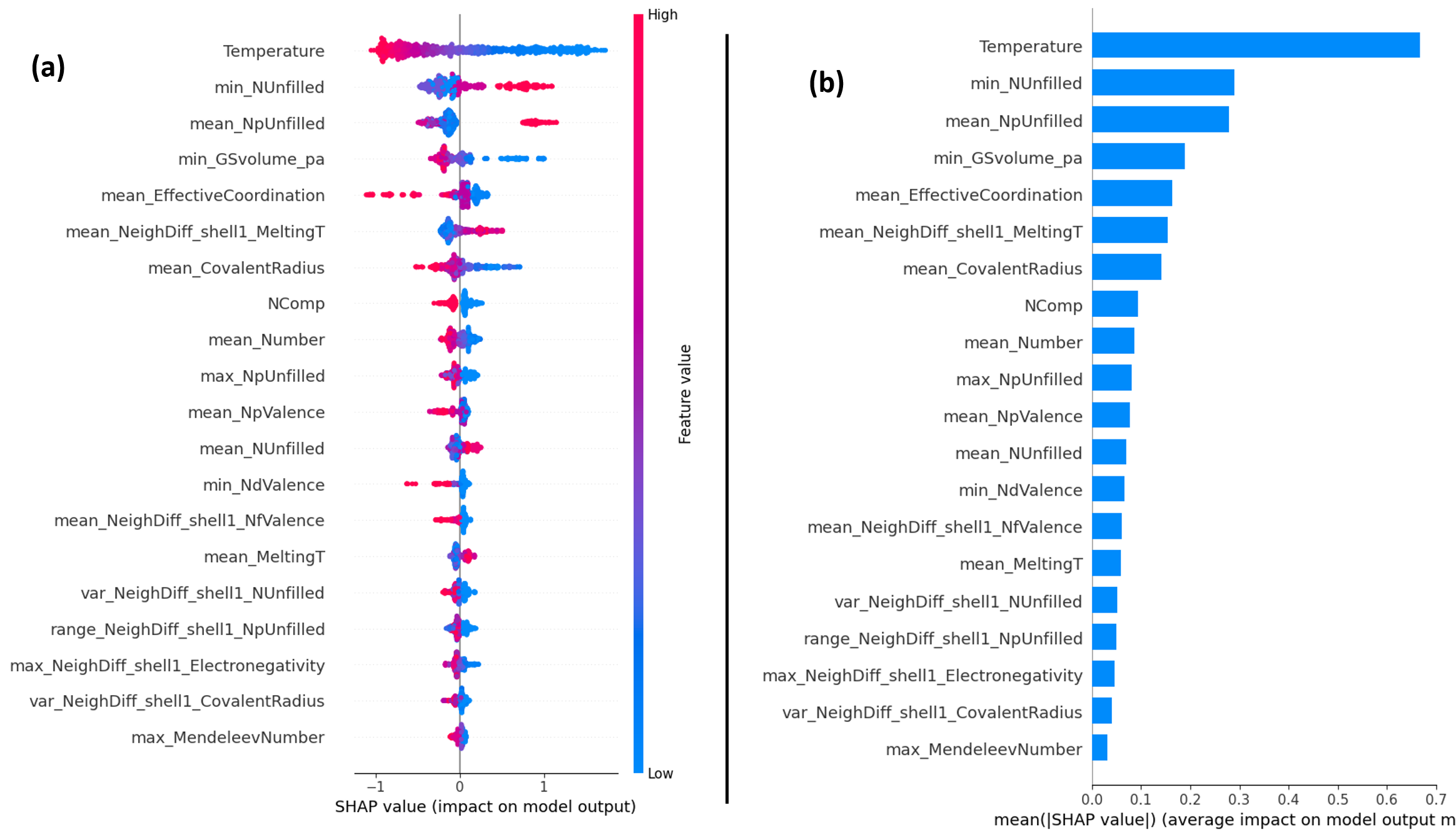}
\par\end{centering}
\caption{\small\justifying\protect\label{SHAP_plots} \textcolor{black}{SHAP (SHapley Additive exPlanations) analysis results of the Extra Trees Regressor model: (a) beeswarm plot illustrating the overall feature influence on the predicted 
$\kappa_L$ values, and (b) feature importance plot showing the relative contribution of individual descriptors.}}
\end{figure*}

To statistically evaluate the model\textquotesingle s predictive performance on the twelve test compounds (low $\kappa_L$ and high $\kappa_L$), we also computed the correlation between the DFT-calculated and ML-predicted values of $\kappa_L$ at various temperatures, as shown in Fig.~\ref{pair_plot}(a). The data points are closely aligned along the ideal line with a slope of 1, indicating a strong linear relationship. The coefficient of determination ($R^2$) for this correlation is 0.961, further supporting the model\textquotesingle s reliability in predicting $\kappa_L$ across different temperature ranges.
In Fig. 9(b), the top 15 most important features used by the Extra Trees Regressor are displayed based on their importance scores. A detailed meaning of each features is described in the SI\cite{supplementary_file}. Among them, temperature(K), the mean number of unfilled p-electrons (mean\_NpUnfilled), minimum number of unfilled electrons (min\_Unfilled) and minimum volume of unitcell (min\_GSvolume\_pa) emerge as the most influential features, with respective importance scores of 0.135, 0.131, 0.106 and 0.073 respectively. The complete list of feature importance score of 53 features are provided in Table S7 of SI\cite{supplementary_file}. These observations underline the strong predictive capability of the model and highlight key physical and chemical descriptors driving accurate predictions of $\kappa_L$. Global Feature importance plot of top 15 features for other trained models is provided in Fig. S3 and S4 of SI\citep{supplementary_file}. It can be observed that for different trained models, the order of important features and their importance score varies, and so do their prediction ability. It is interesting to observer that ETR model performs the best among rest of the models. \textcolor{black}{In the following section, we discuss the SHAP (SHapley Additive exPlanations) analysis\cite{SHAP} explaining why ETR model performed so well.}

\subsection{Physical Interpretation of ETR Model (SHAP analysis)}
\textcolor{black}{Lattice thermal conductivity ($\kappa_L$) can be expressed, within the framework of phonon Boltzmann transport theory, as the sum of contributions from all phonon modes in the Brillouin zone, as follows}

\textcolor{black}{
\begin{equation}
\kappa_L = \frac{1}{V} \sum_{\mathbf{q},s} C_{\mathbf{q},s} \, v_{g}^2(\mathbf{q},s) \, \tau(\mathbf{q},s),
\label{eq:kappa_mode_transport}
\end{equation}}

\textcolor{black}{where $V$ is the crystal volume, $\mathbf{q}$ denotes the phonon wavevector, and $s$ is the phonon branch index. $C_{\mathbf{q},s}$ is the phonon mode heat capacity, $v_{g}(\mathbf{q},s)$ represents the phonon group velocity, and $\tau(\mathbf{q},s)$ is the phonon lifetime (relaxation time). This expression clearly shows that $\kappa_L$ is governed by three key phonon properties: heat capacity, group velocity, and scattering lifetime. Materials with light atoms and stiff bonds typically have higher $v_{g}$, while strong phonon{\em --}phonon scattering or mass disorder reduces $\tau$, leading to lower $\kappa_L$.} 


\textcolor{black}{In order to get insight from model's behavior, SHAP (SHapley Additive exPlanations)\cite{SHAP} analysis is carried out on Extra Tree Regressor(ETR) model. The SHAP beeswarm (Fig.~\ref{SHAP_plots}(a)) and global feature importance (Fig.~\ref{SHAP_plots}(b)) analysis provide clear physical insights into the factors governing lattice thermal conductivity ($\kappa_L$). Temperature has highest influence in controlling $\kappa_L$ where low temperature shows positive SHAP value. Other features associated with unfilled $p$-orbital electrons (mean\_NpUnfilled), covalent radius(mean\_CovalentRadius), and atomic volume (min\_GSvolumn\_pa) exhibit the strongest influence on model predictions. Compounds with higher average unfilled $p$-orbital electrons, smaller atomic volume, and shorter covalent radius show positive SHAP contributions, reflecting higher $\kappa_L$. This is fully consistent with phonon dispersion theory, where lighter and more strongly bonded lattices support higher phonon frequencies and group velocities, thereby enhancing heat transport. In contrast, increased mass contrast, number of elements(NComp), and atomic number(mean\_Number) yield negative SHAP values, indicating reduced $\kappa_L$ due to enhanced phonon scattering arising from mass disorder and structural complexity. Along the same direction, increasing the complexity of crystal structure (mean\_EffectiveCoordination) leads to higher phonon scattering, resulting in reduction in $\kappa_L$ values.} 

\textcolor{black}{Electronic features such as minimum $d$-valence electrons (min\_NdValence) and mean of $p$-orbital valence electrons (mean\_NpValence) also play a significant role. Higher $d$-valence electron (min\_NdValence) content and larger numbers of valence $p$-electrons (mean\_NpValence) correlate with increased lattice anharmonicity and stronger phonon {\em --}phonon scattering, leading to lower $\kappa_L$. Conversely, smaller values of minimum  unfilled electrons are associated with more symmetric bonding and stiffer lattices, resulting in higher $\kappa_L$. Additionally, structural feature such as average of melting point difference corresponding to first coordination shell(mean\_NeighDiff\_shell1\_MeltingT) emerges as a positive predictor, reflecting its correlation with bond stiffness and high Debye temperature. Overall, the Extra Trees Regressor model captures well-established physical trends essential for the thermal transport, validating that its predictive performance arises from meaningful physical correlations rather than spurious statistical patterns.}

\begin{table*}[t]
\caption{\small\justifying\protect\label{tab:table960}Screened compounds from the set of 960 half-Heusler compounds with moderately low $\kappa_L$ (left side) and high $\kappa_L$ (right side) at 300 K}
\centering
\begin{minipage}{.5\textwidth}
    \centering
    \begin{tabular}{l@{\hskip 9mm}c@{\hskip 9mm}r}
    \hline
    Compound & $\kappa_L$($W\,m^{-1}\,K^{-1}$) \\
    \hline
    CsSbBa & 1.423 \\
    CsBiBa & 1.432 \\
    CsBiMg & 1.468 \\
    CsBiSr & 1.469 \\
    CsBiCa & 1.477 \\
    CsBaSb & 1.516 \\
    BaCsSb & 1.523 \\
    CsBaBi & 1.535 \\
    CsBiHg & 1.558 \\
    CsSrBi & 1.564 \\
    CsCaBi & 1.579 \\
    CaCsBi & 1.581 \\
    CsSbSr & 1.597 \\
    MgCsBi & 1.601 \\
    CsAsBa & 1.602 \\
    \hline
    \end{tabular}
    
\end{minipage}%
\begin{minipage}{.5\textwidth}
    \centering
    \begin{tabular}{l@{\hskip 9mm}c@{\hskip 9mm}r}
    \hline
    Compound & $\kappa_L$($W\,m^{-1}\,K^{-1}$) \\
    \hline
    CuBeP     & 20.869 \\
    NaBeN     & 20.280 \\
    LiBeN     & 20.091 \\
    BeCuP     & 20.069 \\
    NaNBe     & 19.593 \\
    LiNMg     & 19.415 \\
    LiNBe     & 19.351 \\
    CuNBe     & 19.275 \\
    CuBeN     & 18.931 \\
    CuPBe     & 18.915 \\
    AgNBe     & 18.910 \\
    BeNaN     & 18.804 \\
    BeLiN     & 18.605 \\
    CuAsBe    & 18.590 \\
    LiNZn     & 18.538 \\
    \hline
    \end{tabular}
\end{minipage}
\end{table*}

\subsection{Material Screening using ML Model}

Once properly trained, machine learning (ML) models can predict material properties much faster than traditional DFT-based calculations. This makes ML models a preferred tool for rapid screening and discovery of novel materials. After validating the Extra Trees Regressor model's performance on unseen compounds from the literature, we applied it to screen materials from a large dataset consisting of 960 Half-Heusler compounds. These compounds were previously examined through high-throughput screening based on their bandgap values and proposed for various renewable energy applications \citep{sahni_reliable_2020}.
The ML model was employed to identify both moderately high and moderately low $\kappa_L$ candidates at room temperature (300 K). The 15 compounds with the lowest and highest predicted $\kappa_L$ values are listed in Table II and Table III, respectively. Among the moderately low $\kappa_L$ candidates, the top three compounds are CsSbBa, CsBiBa, and CsBiMg, with predicted $\kappa_L$ values of 1.423, 1.432, and 1.468 $W\,m^{-1}\,K^{-1}$, respectively. On the other hand, CuBeP, NaBeN, and LiBeN emerged as the top high $\kappa_L$ compounds, with predicted values of 20.869, 20.280, and 20.091 $W\,m^{-1}\,K^{-1}$, respectively.

\begin{table*}[t]
\caption{\small\justifying\protect\label{tab:tableICSD} Selected screened compounds from ICSD database with ultra high $\kappa_L$ (left side) and ultra low $\kappa_L$ (right side) at 300 K.\textcolor{black}{($E_g$ represents bandgap of semiconducting low $\kappa_L$ compounds screened)}}
\centering
\begin{minipage}{.5\textwidth}
    \centering
    \begin{tabular}{l@{\hskip 7mm}l@{\hskip 3mm}c@{\hskip 3mm}r@{\hskip 1mm}r}
    \hline
    Compound & Space group & $\kappa_L$ \textcolor{black}{($W\,m^{-1}\,K^{-1}$)} & \textcolor{black}{ICSD ID}\\
    \hline
    CGe & $P\overline{4}3m$ & 905.99 & 182363 \\
    CSi & $Pn\overline{3}m$ & 591.23 & 182362 \\
    BSb & $P\overline{4}3m$ & 523.25 & 184571 \\
    AsB & $P\overline{4}3m$ & 521.23 & 43871 \\
    BP & $P\overline{4}3m$ & 480.91 & 184570 \\
    AsB$_2$P & $I\overline{4}m2$ & 420.08 & 181293 \\
    BeSe & $P\overline{4}3m$ & 405.19 & 616419 \\
    SiSn & $P\overline{4}3m$ & 392.44 & 184676 \\
    CSn & $P\overline{4}3m$ & 313.25 & 182365 \\
    CTl & $Pn\overline{3}m$ & 305.27 & 618972 \\
    BeTe & $P\overline{4}3m$ & 297.32 & 616439 \\
    C$_4$Si$_2$ & $P4/nmm$ & 286.99 & 187721 \\
    B$_2$N$_2$ & $P6_3/mcm$ & 283.16 & 162870 \\
    BeS & $P\overline{4}3m$ & 237.90 & 186889 \\
    C$_2$Ga$_2$N$_2$Si$_2$ & $Pmn2_1$ & 232.62 & 183047 \\
    C$_2$ & $Fm\overline{3}c$ & 228.90 & 52054 \\
    BBi & $P\overline{4}3m$ & 225.01 & 184569 \\
    C$_{60}$ & $Immm$ & 222.28 & 96620 \\
    C$_3$GaNSi$_3$ & $Pm$ & 220.17 & 183049 \\
    C$_{20}$ & $I\overline{4}3d$ & 217.37 & 185973 \\
    Al$_8$C$_8$Si$_2$ & $P6_3cm$ & 212.55 & 603569 \\
    BeO & $P\overline{4}3m$ & 192.56 & 163821 \\
    Al$_4$C$_3$ & $P\overline{3}c1$ & 187.24 & 52287 \\
    B$_{105}$ & $P\overline{3}c1$ & 187.08 & 14288 \\
    B$_{28}$ & $Pnnm$ & 168.51 & 164659 \\
    \hline
    \end{tabular}
    
\end{minipage}%
\begin{minipage}{.5\textwidth}
    \centering
    \begin{tabular}{l@{\hskip 7mm}l@{\hskip 3mm}c@{\hskip 3mm}r@{\hskip 3mm}r}
    \hline
    Compound & Space group &\ \ \ $\kappa_L$\textcolor{black}{($W\,m^{-1}\,K^{-1}$)} & \textcolor{black}{$E_g$} & \textcolor{black}{ICSD ID} \\
    \hline
    \textbf{CsK$_2$Sb} & $Pn\overline{3}m$ & 0.13 & 0.88 & 53237 \\
    Se$_4$Tl$_3$V & $F\overline{4}3m$ & 0.15 & 1.71 & 652072 \\
    ITl & $I\overline{4}3d$ & 0.16 & 1.79 & 61520 \\
    BiCs$_3$ & $Pn\overline{3}m$ & 0.26 & 0.21 & 659568 \\
    Ag$_4$Ge$_2$S$_6$ & $Cmc2_1$ & 0.31 & 0.99 & 41711 \\
    \textbf{TlBr} & $I\overline{4}3d$ & 0.34 & 1.95 & 181756 \\
    K$_3$Sb & $Pn\overline{3}m$ & 0.38 & 0.68 & 641351 \\
    Rb$_6$Sb$_2$ & $P6_3/mcm$ & 0.42 & 0.10 & 650044 \\
    NbSe$_4$Tl$_3$ & $F\overline{4}3m$ & 0.45 & 2.19 & 600249 \\
    Bi$_4$I$_4$Se$_4$ & $Pnma$ & 0.45 & 1.56 & 280311 \\
    BiK$_3$ & $Pn\overline{3}m$ & 0.48 & 0.18 & 58793 \\
    Se$_4$TaTl$_3$ & $F\overline{4}3m$ & 0.50 & 2.26 & 52431 \\
    Br$_8$Pd$_8$Te$_6$ & $P\overline{1}$ & 0.51 & 0.94 & 418003 \\
    I$_{12}$Pb$_2$Tl$_8$ & $Pbam$ & 0.52 & 2.29 & 100069 \\
    KNa$_2$Sb & $Pn\overline{3}m$ & 0.54 & 0.73 & 44332 \\
    I$_4$Pb$_2$ & $P\overline{3}c1$ & 0.55 & 2.37 & 42014 \\
    Bi$_4$I$_{24}$Tl$_{12}$ & $P2_1/c$ & 0.56 & 2.48 & 417537 \\
    Cs$_{16}$Se$_8$ & $Fdd2$ & 0.56 & 1.81 & 41687 \\
    I$_{20}$Pb$_4$Tl$_{12}$ & $P2_1 2_1 2_1$ & 0.57 & 2.60 & 85309 \\
    AgAsSe$_2$ & $P\overline{3}c1$ & 0.59 & 0.15 & 61708 \\
    I$_{10}$In$_2$Pb$_4$ & $I4/mmm$ & 0.59 & 1.86 & 151999 \\
    Se$_{16}$Ti$_4$Tl$_{16}$ & $P2_1/c$ & 0.61 & 2.20 & 36370 \\
    I$_4$Pd$_2$Te$_2$ & $P\overline{1}$ & 0.63 & 0.67 & 409062 \\
    I$_{12}$Tl$_4$ & $Pnma$ & 0.64 & 1.65 & 61349 \\
    ClTl & $I\overline{4}3d$ & 0.64 & 2.17 & 29107 \\
    \hline
    \end{tabular}
\end{minipage}
\end{table*}

Next, to enable the discovery of novel materials with experimentally verified structures, it is essential to screen compounds from the Inorganic Crystal Structure Database (ICSD) \citep{Zagorac:in5024}. The AFLOW database \citep{aflow}, an open-source repository, provides access to both theoretically simulated and experimentally observed structures from the ICSD. Approximately 60,000 structure files corresponding to ICSD entries were obtained from the AFLOW library. This dataset comprises materials from a wide range of space groups, including experimentally synthesized compounds.
Using our trained Extra Trees Regressor model, these ICSD-based compounds can be rapidly screened for $\kappa_L$ across a broad temperature range (100 \-- 1000 K). The ability to perform high-throughput predictions with the ML model offers a powerful avenue for identifying materials with tailored $\kappa_L$ values, which are essential for diverse applications involving thermal transport and energy conversion.

Table \ref{tab:tableICSD} (left part) presents 20 compounds exhibiting ultra high $\kappa_L$ values exceeding 100 $W\,m^{-1}\,K^{-1}$, screened from the dataset of 60,000 ICSD-based compounds. These high $\kappa_L$ materials are of particular interest for thermal management and heat dissipation technologies.In contrast, Table \ref{tab:tableICSD} (right part) lists 20 compounds identified with ultra low lattice thermal conductivity ($\kappa_L$ < 0.8 $W\,m^{-1}\,K^{-1}$) at 300 K, making them promising candidates for thermoelectric applications.
For example, the ML-predicted high $\kappa_L$ (918.10 $Wm^{-1}K^{-1}$) of CGe aligns well with previously reported first-principles total thermal conductivity value ($\kappa \sim 1200$ $Wm^{-1}K^{-1}$ ) \citep{MUTHAIAH2021100113}, advocating the reliability of our ML-based approach for high $\kappa_L$ predictions. Another significant example is AsB (boron arsenide), where the model\textquotesingle s high predicted $\kappa_L$ (540.93 $Wm^{-1}K^{-1}$) value is corroborated with experimentally measured ultrahigh total $\kappa$ ($\sim$ 1000  $Wm^{-1}K^{-1}$) in AsB single crystals \citep{doi:10.1126/science.aat5522}. It should be noted that,in both the cases, ML predicted lattice thermal conductivity ($\kappa_L$) is compared with the total thermal conductivity($\kappa$), as the DFT computed $\kappa_L$ was unavailable. Other high $\kappa_L$ compounds predicted by the model warrant further theoretical and experimental validation.

On the low $\kappa_L$ side, CsK$_2$Sb was predicted to have a $\kappa_L$ of 0.131 $W\,m^{-1}\,K^{-1}$, in excellent agreement with the first-principles calculations reporting a value of 0.14 $W\,m^{-1}\,K^{-1}$\citep{D2TC03356A}. Similarly, Se$_4$Tl$_3$V is predicted to have a $\kappa_L$ of 0.155 $W\,m^{-1}\,K^{-1}$, closely matching theoretical value of 0.16 $W\,m^{-1}\,K^{-1}$\citep{PhysRevB.102.201201}. The model has also proposed other novel low $\kappa_L$ candidate materials such as BiK$_3$, I$_4$Pb$_2$, Rb$_6$Sb$_2$, and Cs$_8$Te$_4$, which are potential thermoelectric materials and merit further computational and experimental exploration.



Along with high-throughput screening of compounds, we also compared the $\kappa_L$ values predicted by our ML model against experimentally measured values at 300 K, as listed in Table \ref{tab:tableExp}. All experimental $\kappa_L$ data were taken from Ref. \citep{CHEN2019109155}. Notably, the ML model demonstrated a strong ability to predict $\kappa_L$ values close to their experimental counterparts across compounds with diverse space group symmetries (see Table \ref{tab:tableExp}).
It is interesting to note that the model performed well even for compounds whose space group symmetries were underrepresented in the training dataset. For example, the ML predicted $\kappa_L$ of Cr$_2$O$_3$ (space group $R\overline{3}c$) is 15.18 $Wm^{-1}K^{-1}$, closely matching the experimental value of  16.0 $Wm^{-1}K^{-1}$. Similarly, for CoSb$_3$ (space group $Im\overline{3}$), the ML prediction differs by just 0.11 $Wm^{-1}K^{-1}$.
From Table \ref{tab:tableExp}, it is evident that the model shows good agreement with experimental $\kappa_L$ values for both low-symmetry space groups ($Pm\overline{3}m$, $R\overline{3}c$, $I\overline{4}2d$, $Im\overline{3}$, $R\overline{3}m$) and high-symmetry ones ($Fm\overline{3}m$, $F\overline{4}3m$). While the level of agreement varies across individual compounds, the ML model's performance is particularly noteworthy given that it was trained solely on DFT-calculated $\kappa_L$ values. 

\begin{figure*}
\begin{centering}
\includegraphics[scale=0.5]{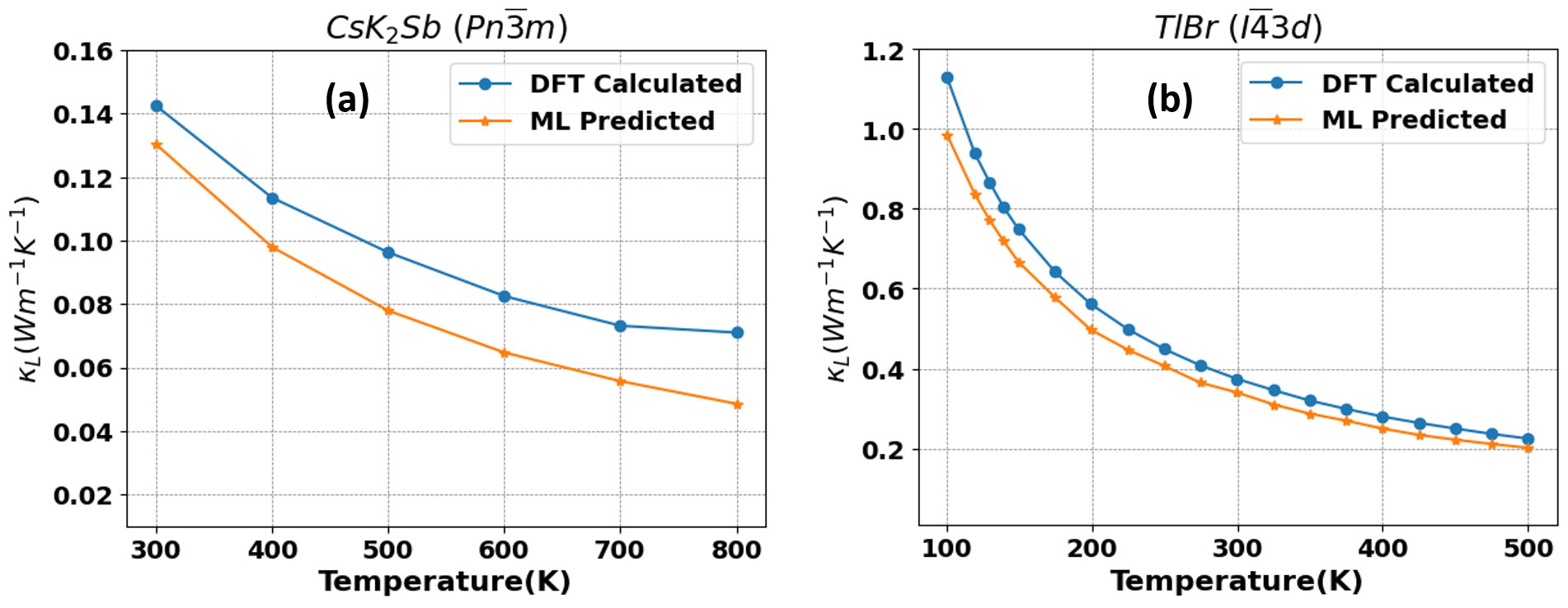}
\par\end{centering}
\caption{\small\justifying\protect\label{fig11}\textcolor{black}{(a,b) Comparison of $\kappa_L$ vs. temperature obtained from ETR model and ab-initio DFT calculations for two compounds, CsK$_2$Sb and TIBr.  }}
\end{figure*}

\textcolor{black}{To further assess the reliability of the machine-learning (ML) screening, we performed a post-screening validation for few low $\kappa_L$ compounds. Since direct first-principles calculations of $\kappa_L$ for the entire screened set are computationally demanding, we adopted a selective validation strategy. A subset of the screened compounds was examined for available theoretical $\kappa_L$ values reported in the literature. Two compounds (CsK$_{2}$Sb $(Pn\overline{3}m)$\citep{D2TC03356A} and TlBr $(I\overline{4}3d)$\citep{PhysRevMaterials.4.045403}) are chosen randomly from the ultra low $\kappa_L$ category in Table 3 and their results obtained from ML model and ab-initio DFT calculations are compared, see Fig.~\ref{fig11}.  It shows remarkable agreement between ML predicted $\kappa_L$ and DFT computed results for the two compounds. More importantly, the model correctly captures the relative trend in $\kappa_L$ across these compounds, with lower $\kappa_L$ observed for materials containing heavier constituent atoms and more complex crystal structures. This agreement indicates that the descriptors employed by the model effectively encode relevant physical factors governing phonon transport. The consistency between ML predictions and literature data for these representative compounds provides additional confidence in the robustness of our screening approach and its applicability for large-scale materials discovery.}

\begin{table*}
\centering
\caption{\small\justifying\protect\label{tab:tableExp} Comparison of experimental $\kappa_L$ \citep{CHEN2019109155} and ML predicted $\kappa_L$ at 300 K.}
\begin{tabular}{|c|c|c|c|c|c|} 
\hline
Sr.No. & Compound & Space Group & $\kappa_{L(exp)}$\textcolor{black}{($W\,m^{-1}\,K^{-1}$)}  & $\kappa_{L(ML)}$\textcolor{black}{($W\,m^{-1}\,K^{-1}$)}  &  $\kappa_{L(exp)}$ - $\kappa_{L(ML)}$\textcolor{black}{($W\,m^{-1}\,K^{-1}$)}  \\
\hline
1 & KCl & $Fm\overline{3}m$ & 7.1 & 5.62 & 1.48 \\
2 & Mg$_2$Sn & $Fm\overline{3}m$ & 7.1 & 5.71 & 1.39 \\
3 & SrTiO$_3$ & $Pm\overline{3}m$ & 8.5 & 7.12 & 1.38 \\
4 & GaCuTe$_2$ & $I\overline{4}2d$ & 2.2 & 0.97 & 1.23 \\
5 & Cr$_2$O$_3$ & $R\overline{3}c$ & 16.0 & 15.18 & 0.82 \\
6 & RbBr & $Fm\overline{3}m$ & 3.8 & 3.15 & 0.65 \\
7 & PbSe & $Fm\overline{3}m$ & 2.0 & 1.47 & 0.53 \\
8 & GaAgS$_2$ & $I\overline{4}2d$ & 1.45 & 1.29 & 0.16 \\
9 & NaCl & $Fm\overline{3}m$ & 7.1 & 7.10 & -0.00 \\
10 & CoSb$_3$ & $Im\overline{3}$ & 10.0 & 10.11 & -0.11 \\
11 & Sb$_2$Te$_3$ & $R\overline{3}m$ & 2.4 & 2.72 & -0.32 \\
12 & TePb & $Fm\overline{3}m$ & 2.5 & 3.00 & -0.50 \\
13 & RbI & $Fm\overline{3}m$ & 2.3 & 2.92 & -0.62 \\
14 & LiF & $Fm\overline{3}m$ & 17.6 & 18.22 & -0.62 \\
15 & KI & $Fm\overline{3}m$ & 2.6 & 3.37 & -0.77 \\
16 & Bi$_2$Te$_3$ & $R\overline{3}m$ & 1.6 & 2.40 & -0.80 \\
17 & Bi$_2$Se$_3$ & $R\overline{3}m$ & 1.34 & 2.15 & -0.81 \\
18 & KBr & $Fm\overline{3}m$ & 3.4 & 4.48 & -1.08 \\
19 & CsI & $Fm\overline{3}m$ & 1.1 & 2.30 & -1.20 \\
20 & CdTe & $F\overline{4}3m$ & 7.5 & 8.92 & -1.42 \\
\hline
\end{tabular}
\end{table*}

\textcolor{black}{It is worth noting that in several cases (see Fig.~\ref{fig7}, and Table 4), the ML predicted $\kappa_L$ values are lower than the experimental counterparts, which appears opposite to the conventional expectation from ab-initio phonon-based calculations. Indeed, in conventional ab-initio phonon-based calculations, the computed lattice thermal conductivity ($\kappa_L$) often overestimates the experimental values because real samples inevitably contain impurities, defects, and grain boundaries that act as additional phonon-scattering centers. However, there exists few cases where theoretical $\kappa_L$ value is underestimated. The underestimation of simulated $\kappa_L$ relative to experiment arises mainly due to two factors, (1) The majority of training data are from low-to-moderate $\kappa_L$ compounds, where DFT itself is known to sometimes overestimate phonon scattering (due to approximate treatment of anharmonicity or neglect of isotope effects). Consequently, the ML model learns a slightly suppressed $\kappa_L$ scale (2) In several cases (e.g.,  Cu$_3$VSe$_4$\cite{paper1}, Cu$_2$O\cite{paper2}, U$_3$H$_8$\cite{paper3}, TlXTe$_2$\cite{paper4} (X=Ga,In)), the experimental $\kappa_L$ values reported in the literature are higher due to sample quality, stoichiometric control, or measurement uncertainties. Since our ML predictions are based on generalized statistical patterns from diverse datasets, they may not capture such system-specific extrinsic enhancements.
However, one should note that this trend is not uniform — for a number of materials e.g., Bi$_2$Te$_3$, Sb$_2$Te$_3$, the ML predictions are higher than experiment, as expected. Thus, the small systematic underestimation seen in Table 4 should be viewed as a reflection of (i) the mixed nature of DFT references used in training, and (ii) the averaging inherent in machine learning, which may not reproduce system-specific outliers exactly.}

\section{Conclusion}

Lattice thermal conductivity ($\kappa_L$) is a key parameter in identifying high-performance thermoelectric materials. However, ab initio calculations of $\kappa_L$, particularly those based on density functional theory (DFT) combined with phonon calculations, are among the most computationally intensive tasks in materials science. This challenge motivates the development of efficient surrogate methods that can significantly reduce computational overhead, yet maintaining the similar level of accuracy.
In this work, we present a machine learning (ML) model designed to predict $\kappa_L$ with DFT-level accuracy, enabling rapid and reliable screening of low-$\kappa_L$ materials without the need for time-consuming phonon computations. We employ the Extra Trees Regressor (ETR) algorithm to construct the model, which demonstrates strong predictive performance across a broad temperature range. The model\textquotesingle s robustness and generalizability are validated on a diverse set of low-symmetry compounds, capturing temperature-dependent trends in $\kappa_L$ with high fidelity.
Our results show that the ML model effectively identifies materials with ultralow $\kappa_L$, closely matching DFT predictions and accurately reflecting the upper bounds of $\kappa_L$ values. We apply the trained model to screen 960 in-house half-Heusler compounds and approximately 60,000 materials from the ICSD database curated through the AFLOW library, uncovering several promising candidates with both ultralow and ultrahigh $\kappa_L$ for further theoretical investigation. \textcolor{black}{We have also validated ML predicted $\kappa_L$ of two screened promising compounds with their corresponding DFT computed results.}
Finally, we benchmark the model's predictions against available experimental data, confirming its reliability and practical utility. A key advantage of this approach is the dramatic reduction in computational time from several weeks per material using ab initio methods to milliseconds using ML thus enabling efficient high-throughput discovery of novel thermoelectric materials.

\section*{Acknowledgments}
\textcolor{black}{The authors acknowledge financial support from the Centre for Machine Learning and Data Science (CMInDS), Indian Institute of Technology(IIT) Bombay, through the research project funding, 
and appreciates access to the computational facilities and research infrastructure utilized in this study.}

\section*{Code and Data Availability}
\textcolor{black}{The lattice thermal conductivity ($\kappa_L$) values used in this study were extracted from previously published literature sources, as cited in Table S1 and S2, using digital data extraction methods. These values were not recalculated from first principles. The compiled dataset and the machine learning scripts used for model development and analysis are available from the corresponding author upon reasonable request.}

\bibliographystyle{unsrt}
\bibliography{Ref}

\end{document}


\title{Supplementary Information for {\it ``Accelerated Prediction of Temperature-Dependent Lattice Thermal Conductivity via Ensembled Machine Learning Models"}}
\author{Piyush Paliwal$^1$}
\author{Aftab Alam$^{1,2}$}
\email{aftab@iitb.ac.in}
\affiliation{$^1$Center for Machine Intelligence and Data Science (CMInDS), Indian Institute of Technology Bombay, Powai, Mumbai 400076, India}
\affiliation{$^2$Department of Physics, Indian Institute of Technology Bombay, Powai, Mumbai 400076, India}

\maketitle

Here, we provide further auxiliary details regarding those compounds used for trainig, compositional and structural features, feature importances and formulation of all the models.

\section{Compound data for Model Training}

\begin{table}[H]
\caption{Data for 116 Compounds with their corresponding Space Group\citep{Juneja2019}. }
\label{table:compdata1}
\begin{tabularx}{0.95\textwidth} { 
  | >{\centering\arraybackslash}X 
  | >{\centering\arraybackslash}X 
  | >{\centering\arraybackslash}X
  | >{\centering\arraybackslash}X
  | >{\centering\arraybackslash}X
  | >{\centering\arraybackslash}X | }
\hline
\textbf{Sr. No.} & \textbf{Compound Name} & \textbf{Space Group} & \textbf{Sr. No.} & \textbf{Compound Name} & \textbf{Space Group} \\ 
\hline
1 & AlAs & $F\overline{4}3m$ & 59 & HfSe$_2$ & $P\overline{3}m1$ \\ 
2 & AlP & $F\overline{4}3m$ & 60 & HfSnPd & $F\overline{4}3m$ \\ 
3 & AlSb & $F\overline{4}3m$ & 61 & HoHSe & $P\overline{6}m2$ \\ 
4 & BP & $F\overline{4}3m$ & 62 & HoNiSb & $F\overline{4}3m$ \\ 
5 & BSb & $F\overline{4}3m$ & 63 & HoSbPd & $F\overline{4}3m$ \\ 
6 & BaS & $F\overline{4}3m$ & 64 & HoSbPt & $F\overline{4}3m$ \\ 
7 & BaSe & $Fm\overline{3}m$ & 65 & HoSnAu & $F\overline{4}3m$ \\ 
8 & BaTe & $Fm\overline{3}m$ & 66 & K$_2$S & $Fm\overline{3}m$ \\ 
9 & BeS & $F\overline{4}3m$ & 67 & K$_2$Se & $Fm\overline{3}m$ \\ 
10 & BeSe & $F\overline{4}3m$ & 68 & K$_2$Te & $Fm\overline{3}m$ \\ 
11 & BeTe & $F\overline{4}3m$ & 69 & LaCoTe & $F\overline{4}3m$ \\ 
12 & BiB & $F\overline{4}3m$ & 70 & Li$_2$S & $Fm\overline{3}m$ \\ 
13 & CaS & $Fm\overline{3}m$ & 71 & Li$_2$Se & $Fm\overline{3}m$ \\ 
14 & CaSe & $F\overline{4}3m$ & 72 & Li$_2$Te & $Fm\overline{3}m$ \\ 
15 & CaSe & $Fm\overline{3}m$ & 73 & LiFeP & $I4mm$ \\ 
16 & CaTe & $Fm\overline{3}m$ & 74 & LiGaSi & $F\overline{4}3m$ \\ 
17 & CdS & $F\overline{4}3m$ & 75 & LiMgP & $F\overline{4}3m$ \\ 
18 & CdSe & $F\overline{4}3m$ & 76 & LuSbPt & $F\overline{4}3m$ \\ 
19 & CdTe & $F\overline{4}3m$ & 77 & MoS$_2$ & $P\overline{6}m2$ \\ 
20 & GaAs & $F\overline{4}3m$ & 78 & MoSe$_2$ & $P\overline{6}m2$ \\ 
21 & GaP & $F\overline{4}3m$ & 79 & MoSe$_2$ & $R\overline{3}m$ \\ 
22 & GeC & $F\overline{4}3m$ & 80 & Na$_2$S & $Fm\overline{3}m$ \\ 
23 & InP & $F\overline{4}3m$ & 81 & Na$_2$Se & $Fm\overline{3}m$ \\ 
24 & MgS & $F\overline{4}3m$ & 82 & Na$_2$Te & $Fm\overline{3}m$ \\ 
25 & MgS & $Fm\overline{3}m$ & 83 & NbCoSn & $F\overline{4}3m$ \\
\hline
\end{tabularx}
\end{table}

\begin{table}[H]
\label{table:compdata2}
\begin{tabularx}{0.95\textwidth} { 
  | >{\centering\arraybackslash}X 
  | >{\centering\arraybackslash}X 
  | >{\centering\arraybackslash}X
  | >{\centering\arraybackslash}X
  | >{\centering\arraybackslash}X
  | >{\centering\arraybackslash}X | }

\hline
\textbf{Sr. No.} & \textbf{Compound Name} & \textbf{Space Group} & \textbf{Sr. No.} & \textbf{Compound Name} & \textbf{Space Group} \\ 
\hline
26 & MgSe & $F\overline{4}3m$ & 84 & NbFeSb & $F\overline{4}3m$ \\ 
27 & MgTe & $Fm\overline{3}m$ & 85 & NbSbRu & $F\overline{4}3m$ \\ 
28 & MgTe & $F\overline{4}3m$ & 86 & PbF$_2$ & $Fm\overline{3}m$ \\ 
29 & PbS & $Fm\overline{3}m$ & 87 & PbI$_2$ & $R\overline{3}m$ \\ 
30 & PbSe & $Fm\overline{3}m$ & 88 & PbI$_2$ & $P\overline{3}m1$ \\
31 & SiC & $F\overline{4}3m$ & 89 & PtS$_2$ & $P\overline{3}m1$ \\ 
32 & SiSn & $F\overline{4}3m$ & 90 & Rb$_2$S & $Fm\overline{3}m$ \\ 
33 & SnC & $F\overline{4}3m$ & 91 & Rb$_2$Se & $Fm\overline{3}m$ \\ 
34 & SnS & $Fm\overline{3}m$ & 92 & Rb$_2$Te & $Fm\overline{3}m$ \\ 
35 & SnSe & $Fm\overline{3}m$ & 93 & ScCoTe & $F\overline{4}3m$ \\ 
36 & SrTe & $Fm\overline{3}m$ & 94 & ScSnAu & $F\overline{4}3m$ \\ 
37 & TePb & $Fm\overline{3}m$ & 95 & SmSnAu & $F\overline{4}3m$ \\ 
38 & TlBr & $Pm\overline{3}m$ & 96 & SnO$_2$ & $R\overline{3}m$ \\ 
39 & TlBr & $Fm\overline{3}m$ & 97 & SeTePd & $F\overline{4}3m$ \\ 
40 & TlCl & $Pm\overline{3}m$ & 98 & TaTlPt & $F\overline{4}3m$ \\ 
41 & TlCl & $Fm\overline{3}m$ & 99 & Te$_2$Mo & $P\overline{6}m2$ \\ 
42 & TlI & $Pm\overline{3}m$ & 100 & AlP & $P6_3mc$ \\ 
43 & BaLiP & $P\overline{6}m2$ & 101 & B$_2$AsP & $P\overline{4}m2$ \\ 
44 & BiTeI & $P3m1$ & 102 & BP & $P6_3mc$ \\ 
45 & Ca$_2$Ge & $Fm\overline{3}m$ & 103 & Ba$_2$AgSb & $Fm\overline{3}m$ \\ 
46 & Ca$_2$Pb & $Fm\overline{3}m$ & 104 & Ba$_2$BiAu & $Fm\overline{3}m$ \\ 
47 & Ca$_2$Si & $Fm\overline{3}m$ & 105 & Ba$_2$PCl & $R\overline{3}m$ \\ 
48 & Ca$_2$Sn & $Fm\overline{3}m$ & 106 & Ba$_2$SbAu & $Fm\overline{3}m$ \\ 
49 & CeSe$_2$ & $Fm\overline{3}m$ & 107 & CdS & $P6_3mc$ \\ 
50 & Cs$_2$Se & $Fm\overline{3}m$ & 108 & CdSe & $P6_3mc$ \\ 
51 & DySbPt & $F\overline{4}3m$ & 109 & CdTe & $P6_3mc$ \\ 
\hline
\end{tabularx}
\end{table}

\begin{table}[H]
\label{table:compdata3}
\begin{tabularx}{0.95\textwidth} { 
  | >{\centering\arraybackslash}X 
  | >{\centering\arraybackslash}X 
  | >{\centering\arraybackslash}X
  | >{\centering\arraybackslash}X
  | >{\centering\arraybackslash}X
  | >{\centering\arraybackslash}X | }

\hline
\textbf{Sr. No.} & \textbf{Compound Name} & \textbf{Space Group} & \textbf{Sr. No.} & \textbf{Compound Name} & \textbf{Space Group} \\ 
\hline
52 & ErNiSb & $F\overline{4}3m$ & 110 & CsHoS$_2$ & $R\overline{3}m$ \\ 
53 & ErSbPd & $F\overline{4}3m$ & 111 & CsK$_2$Sb & $Fm\overline{3}m$ \\ 
54 & ErSbPt & $F\overline{4}3m$ & 112 & DyTlS$_2$ & $R\overline{3}m$ \\ 
55 & ErSnAu & $F\overline{4}3m$ & 113 & DyTlSe$_2$ & $R\overline{3}m$ \\ 
56 & FeS$_2$ & $R\overline{3}m$ & 114 & DyTlTe$_2$ & $R\overline{3}m$ \\ 
57 & HfNiSn & $F\overline{4}3m$ & 115 & GaAs & $P6_3mc$ \\ 
58 & HfS$_2$ & $P\overline{3}m1$ & 116 & GaP & $P6_3mc$ \\ 
\hline
\end{tabularx}
\end{table}

\begin{table}[H]
\caption{Data for the rest of the compounds (117 to 150) in the training data-set and their corresponding Space Group. }
\label{table:compdata4}
\begin{tabularx}{0.8\textwidth} { 
  | >{\centering\arraybackslash}X 
  | >{\centering\arraybackslash}X 
  | >{\centering\arraybackslash}X
  | >{\centering\arraybackslash}X | }
\hline
\textbf{Sr. No.} & \textbf{Compound Name} & \textbf{Space Group} & \textbf{Ref.} \\ 
\hline
117 & Ag$_2$GeTe$_3$ & $Cmc2_1$ & \citep{FanMo2024}\\ 
118 & Ag$_2$GeS$_3$ & $Cmc2_1$ & \citep{FanMo2024}\\ 
119 & Ag$_2$GeSe$_3$ & $Cmc2_1$ & \citep{FanMo2024}\\ 
120 & BeSiSb$_2$ & $I\overline{4}2d$ & \citep{GOVINDARAJ2023}\\ 
121 & BiSeI & $Pnma$ & \citep{HE2022}\\ 
122 & CoNiFe & $F\overline{4}3m$ & \citep{zhang2024}\\ 
123 & CsZnSb & $P6_3/mmc$ & \citep{haque2021}\\ 
124 & Li$_2$CaPb & $Fm\overline{3}m$ & \citep{SHARMA2023106289}\\ 
125 & RbZnSb & $P6_3/mmc$ & \citep{haque2021}\\ 
126 & MgSiSb$_2$ & $I\overline{4}2d$ & \citep{GOVINDARAJ2023}\\ 
127 & Na$_2$MgSn & $P6_3/mmc$ & \citep{wang2019}\\ 
128 & Na$_2$MgPb & $P6_3/mmc$ & \citep{wang2019}\\ 
129 & CaP$_3$ & $P\overline{1}$ & \citep{zhu_significant_2020}\\ 
130 & NiHfSn & $F\overline{4}3m$ & \citep{Andrea_2015}\\  
131 & NiTiSn & $F\overline{4}3m$ & \citep{Andrea_2015}\\ 
132 & RhBiHf & $F\overline{4}3m$ & \citep{wei2023}\\ 
133 & RhBiTi & $F\overline{4}3m$ & \citep{wei2023}\\ 
134 & Sr$_2$Ge & $Pnma$ & \citep{YOU2023101015}\\ 
135 & ScCuTe$_2$ & $P3m1$ & \citep{RUGUT2021}\\ 
136 & ScCuSe$_2$ & $P3m1$ & \citep{RUGUT2021}\\ 
137 & ScCuS$_2$ & $P3m1$ & \citep{RUGUT2021}\\ 
138 & Tl$_3$VSe$_4$ & $I\overline{4}3m$ & \citep{YANG2023}\\ 
139 & Cu$_3$VSe$_4$ & $P\overline{4}3m$ & \citep{YANG2023}\\ 
140 & CuBiTeO & $P4/nmm$ & \citep{CuBiTeO}\\ 
141 & ZrCoSb & $F\overline{4}3m$ & \citep{Gandi2017}\\ 
\hline
\end{tabularx}
\end{table}

\begin{table}[H]
\label{table:compdata5}
\begin{tabularx}{0.8\textwidth} { 
  | >{\centering\arraybackslash}X 
  | >{\centering\arraybackslash}X 
  | >{\centering\arraybackslash}X
  | >{\centering\arraybackslash}X | }
\hline
\textbf{Sr. No.} & \textbf{Compound Name} & \textbf{Space Group} & \textbf{Ref.} \\ 
\hline
142 & HfCoSb & $F\overline{4}3m$ & \citep{Gandi2017}\\ 
143 & LiMgN & $F\overline{4}3m$ & \citep{LUO2023}\\ 
144 & LiZnN & $F\overline{4}3m$ & \citep{LUO2023}\\ 
145 & ScRhTe & $F\overline{4}3m$ & \citep{SINGH2017120}\\ 
146 & LiScPtGe & $F\overline{4}3m$ & \citep{JASPAL2023161}\\ 
147 & LiYPtSn & $F\overline{4}3m$ & \citep{JASPAL2023161}\\ 
148 & LiYPdPb & $F\overline{4}3m$ & \citep{JASPAL2023161}\\ 
149 & BaAgP & $P6_3/mmc$ & \citep{PARVIN2021104068}\\  
150 & Sr$_2$Si & $Pnma$ & \citep{YOU2023101015}\\  
\hline
\end{tabularx}
\end{table}

\textcolor{black}{\textbf{Target Property: Lattice Thermal Conductivity ($\kappa_L$)}: 
The lattice thermal conductivity ($\kappa_L$) values (between 100K to 1200K) for all 150 compounds listed in Table \ref{table:compdata1} and Table \ref{table:compdata4} were obtained from density functional theory (DFT) calculations performed at the Perdew–Burke–Ernzerhof (PBE) level of the generalized gradient approximation (GGA). The phonon properties were computed by explicitly considering third-order interatomic force constants (IFCs), and the lattice thermal conductivities were subsequently evaluated by solving the phonon Boltzmann transport equation (BTE). These $\kappa_L$ values were used as the reference dataset for training and validating the machine learning models presented in this study.}

\begin{figure*}[t]
     \centering
      \includegraphics[width=0.95\linewidth]{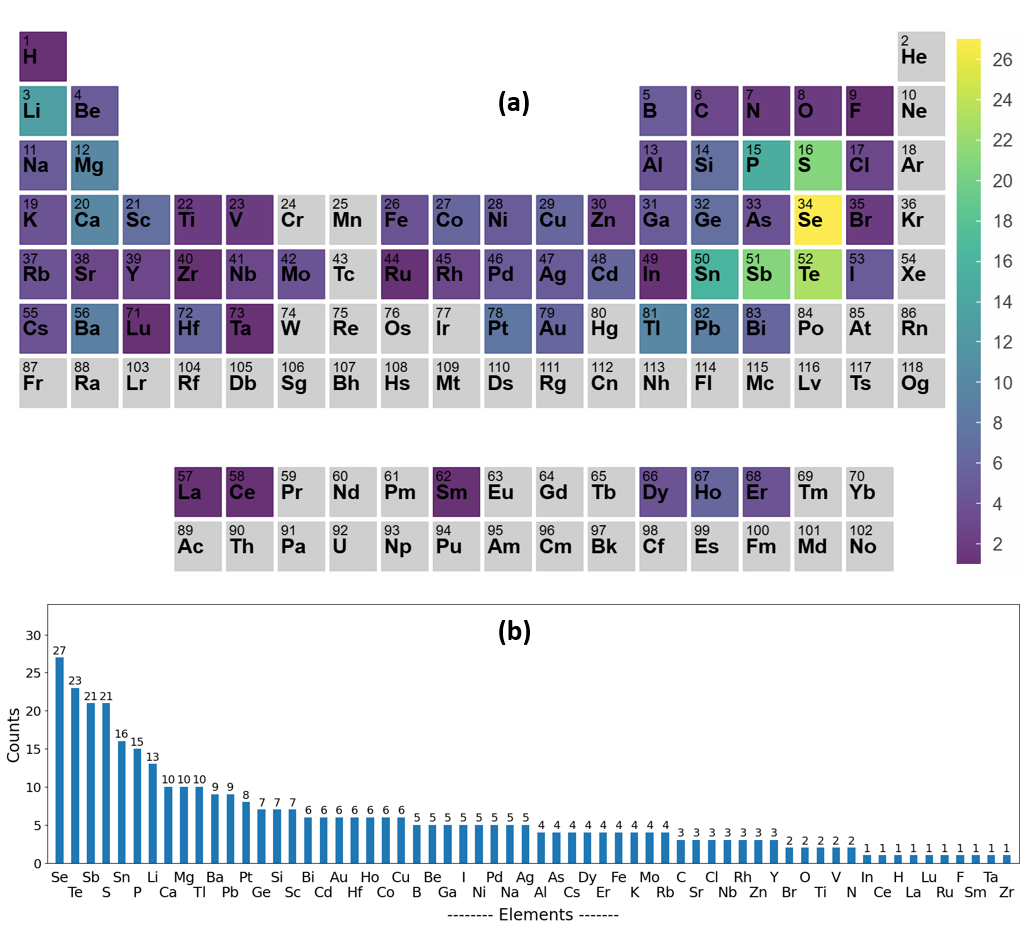}
      \caption{\small \justifying \protect\label{figures:figure_S1.png} \textcolor{black}{The distribution of elements across the dataset of 150 compounds in (a) a periodic table map (color bar indicates the frequency of occurrence of each element) (b) a histogram representing a complementary quantitative overview.}}
        \label{figure-1}
\end{figure*}

\clearpage
\section{Compositional and Structure features}
The features used in this work were obtained from the Material Agonistic Platform for Informatics and Exploration (Magpie)\citep{magpie_code}. In broad category, feature space is divided mainly into two categories: Structural characteristics (126 in number) and composition characteristics (145 in number). This results in a total of 271 features corresponding to each crystal structure. Both kinds of features are generated from VASP formatted structure file (namely POSCAR) of the concerned compound. A brief description of each feature category is provided below. For detailed information, refer to the paper by Ward et al. and its Supplementary Section\citep{Ward2016}.    

\subsection{Composition Features}
Table \ref{tab6} display the elemental properties used to generate compositional features. Table \ref{tab7} provides 145 composition features generated using 6 equations of mean, mean absolute deviation, range, maximum, minimum and mode fraction of each element in the crystal collectively with its 22 properties as provided in Table \ref{tab6}.

\begin{table}[t]
\caption{ 22 Elemental properties used to generate the composition features.}
\begin{tabularx}{1.0\textwidth} { 
  | >{\centering\arraybackslash}X 
  | >{\centering\arraybackslash}X 
   | }
\hline
\textbf{Elemental property} & \textbf{Elemental property} \\ 
\hline

Atomic number & s valance electrons \\ 
Atomic weight & p valance electrons \\ 
Electronegativity & d valance electrons \\ 
Bandgap of ground state & f valance electrons \\ 
Space group number of ground state & Total valance  electrons \\ 
Mendeleev number & Unfilled s states \\
Column & Unfilled p states \\
Row & Unfilled d states \\
Column radius & Unfilled f states \\
Specific volume of ground state & Total unfilled states \\
Magnetic moment of ground state & Melting temperature \\
\hline
\end{tabularx}
\label{tab6}
\end{table}


\begin{table}[H]
\centering
\caption{Complete list of Composition features. }
\begin{tabular}{|c|l|c|l|}
\hline
\textbf{Sr. No.} & \textbf{Feature Name} & \textbf{Sr. No.} & \textbf{Feature Name} \\
\hline
1 & NComp & 74 & maxdiff\_NfValence \\
2 & Comp\_L2Norm & 75 & dev\_NfValence \\
3 & Comp\_L3Norm & 76 & max\_NfValence \\
4 & Comp\_L5Norm & 77 & min\_NfValence \\
5 & Comp\_L7Norm & 78 & most\_NfValence \\
6 & Comp\_L10Norm & 79 & mean\_NValance \\
7 & mean\_Number & 80 & maxdiff\_NValance \\
8 & maxdiff\_Number & 81 & dev\_NValance \\
9 & dev\_Number & 82 & max\_NValance \\
10 & max\_Number & 83 & min\_NValance \\
11 & min\_Number & 84 & most\_NValance \\
12 & most\_Number & 85 & mean\_NsUnfilled \\
13 & mean\_MendeleevNumber & 86 & maxdiff\_NsUnfilled \\
14 & maxdiff\_MendeleevNumber & 87 & dev\_NsUnfilled \\
15 & dev\_MendeleevNumber & 88 & max\_NsUnfilled \\
16 & max\_MendeleevNumber & 89 & min\_NsUnfilled \\
17 & min\_MendeleevNumber & 90 & most\_NsUnfilled \\
18 & most\_MendeleevNumber & 91 & mean\_NpUnfilled \\
19 & mean\_AtomicWeight & 92 & maxdiff\_NpUnfilled \\
20 & maxdiff\_AtomicWeight & 93 & dev\_NpUnfilled \\
21 & dev\_AtomicWeight & 94 & max\_NpUnfilled \\
22 & max\_AtomicWeight & 95 & min\_NpUnfilled \\
23 & min\_AtomicWeight & 96 & most\_NpUnfilled \\
24 & most\_AtomicWeight & 97 & mean\_NdUnfilled \\
25 & mean\_MeltingT & 98 & maxdiff\_NdUnfilled \\
26 & maxdiff\_MeltingT & 99 & dev\_NdUnfilled \\
27 & dev\_MeltingT & 100 & max\_NdUnfilled \\

\hline
\end{tabular}
\label{tab7}
\end{table}

\begin{table}[H]
\centering
\begin{tabular}{|c|l|c|l|}
\hline
\textbf{Sr. No.} & \textbf{Feature Name} & \textbf{Sr. No.} & \textbf{Feature Name} \\
\hline
28 & max\_MeltingT & 101 & min\_NdUnfilled \\
29 & min\_MeltingT & 102 & most\_NdUnfilled \\
30 & most\_MeltingT & 103 & mean\_NfUnfilled \\
31 & mean\_Column & 104 & maxdiff\_NfUnfilled \\
32 & maxdiff\_Column & 105 & dev\_NfUnfilled \\
33 & dev\_Column & 106 & max\_NfUnfilled \\
34 & max\_Column & 107 & min\_NfUnfilled \\
35 & min\_Column & 108 & most\_NfUnfilled \\
36 & most\_Column & 109 & mean\_NUnfilled \\
37 & mean\_Row & 110 & maxdiff\_NUnfilled \\
38 & maxdiff\_Row & 111 & dev\_NUnfilled \\
39 & dev\_Row & 112 & max\_NUnfilled \\
40 & max\_Row & 113 & min\_NUnfilled \\
41 & min\_Row & 114 & most\_NUnfilled \\
42 & most\_Row & 115 & mean\_GSvolume\_pa \\
43 & mean\_CovalentRadius & 116 & maxdiff\_GSvolume\_pa \\
44 & maxdiff\_CovalentRadius & 117 & dev\_GSvolume\_pa \\
45 & dev\_CovalentRadius & 118 & max\_GSvolume\_pa \\
46 & max\_CovalentRadius & 119 & min\_GSvolume\_pa \\
47 & min\_CovalentRadius & 120 & most\_GSvolume\_pa \\
48 & most\_CovalentRadius & 121 & mean\_GSbandgap \\
49 & mean\_Electronegativity & 122 & maxdiff\_GSbandgap \\
50 & maxdiff\_Electronegativity & 123 & dev\_GSbandgap \\
51 & dev\_Electronegativity & 124 & max\_GSbandgap \\
52 & max\_Electronegativity & 125 & min\_GSbandgap \\
53 & min\_Electronegativity & 126 & most\_GSbandgap \\
54 & most\_Electronegativity & 127 & mean\_GSmagmom \\
55 & mean\_NsValence & 128 & maxdiff\_GSmagmom \\

\hline
\end{tabular}
\end{table}

\begin{table}[htbp]
\centering
\begin{tabular}{|c|l|c|l|}
\hline
\textbf{Sr. No.} & \textbf{Feature Name} & \textbf{Sr. No.} & \textbf{Feature Name} \\
\hline
56 & maxdiff\_NsValence & 129 & dev\_GSmagmom \\
57 & dev\_NsValence & 130 & max\_GSmagmom \\
58 & max\_NsValence & 131 & min\_GSmagmom \\
59 & min\_NsValence & 132 & most\_GSmagmom \\
60 & most\_NsValence & 133 & mean\_SpaceGroupNumber \\
61 & mean\_NpValence & 134 & maxdiff\_SpaceGroupNumber \\
62 & maxdiff\_NpValence & 135 & dev\_SpaceGroupNumber \\
63 & dev\_NpValence & 136 & max\_SpaceGroupNumber \\
64 & max\_NpValence & 137 & min\_SpaceGroupNumber \\
65 & min\_NpValence & 138 & most\_SpaceGroupNumber \\
66 & most\_NpValence & 139 & frac\_sValence \\
67 & mean\_NdValence & 140 & frac\_pValence \\
68 & maxdiff\_NdValence & 141 & frac\_dValence \\
69 & dev\_NdValence & 142 & frac\_fValence \\
70 & max\_NdValence & 143 & CanFormIonic \\
71 & min\_NdValence & 144 & MaxIonicChar \\
72 & most\_NdValence & 145 & MeanIonicChar \\
73 & mean\_NfValence & & \\
\hline
\end{tabular}
\end{table}

\newpage
\subsection{Structural Features}
Table \ref{tab9} provides 126 structural features. These features are generated using mean, maximum, variance, and range of various structural properties.

\FloatBarrier
\begin{table}[H]
\centering
\caption{Complete list of structural features. }
\begin{tabular}{|c|l|c|l|}
\hline
Sr. No. & Feature Name & Sr. No. & Feature Name \\
\hline
1 & max\_NeighDiff\_shell1\_GSvolume\_pa & 64 & mean\_NeighDiff\_shell1\_NfUnfilled \\
2 & mean\_WCMagnitude\_Shell2 & 65 & min\_NeighDiff\_shell1\_MeltingT \\
3 & mean\_WCMagnitude\_Shell1 & 66 & max\_NeighDiff\_shell1\_NValance \\
4 & mean\_NeighDiff\_shell1\_Row & 67 & min\_NeighDiff\_shell1\_CovalentRadius \\
5 & var\_BondLengthVariation & 68 & range\_NeighDiff\_shell1\_Electronegativity \\
6 & var\_NeighDiff\_shell1\_SpaceGroupNumber & 69 & var\_EffectiveCoordination \\
7 & range\_NeighDiff\_shell1\_SpaceGroupNumber & 70 & range\_NeighDiff\_shell1\_NsUnfilled \\
8 & range\_NeighDiff\_shell1\_AtomicWeight & 71 & max\_NeighDiff\_shell1\_NsValence \\
9 & max\_NeighDiff\_shell1\_Column & 72 & min\_NeighDiff\_shell1\_NUnfilled \\
10 & max\_NeighDiff\_shell1\_GSmagmom & 73 & min\_NeighDiff\_shell1\_NpValence \\
11 & min\_NeighDiff\_shell1\_NfUnfilled & 74 & var\_NeighDiff\_shell1\_NfUnfilled \\
12 & mean\_NeighDiff\_shell1\_AtomicWeight & 75 & mean\_WCMagnitude\_Shell3 \\
13 & max\_NeighDiff\_shell1\_MendeleevNumber & 76 & mean\_NeighDiff\_shell1\_NfValence \\
14 & range\_NeighDiff\_shell1\_Row & 77 & mean\_NeighDiff\_shell1\_NdUnfilled \\
15 & range\_NeighDiff\_shell1\_NpValence & 78 & mean\_NeighDiff\_shell1\_SpaceGroupNumber \\
16 & min\_NeighDiff\_shell1\_SpaceGroupNumber & 79 & max\_NeighDiff\_shell1\_NdUnfilled \\
17 & var\_NeighDiff\_shell1\_GSbandgap & 80 & max\_NeighDiff\_shell1\_GSbandgap \\
18 & min\_NeighDiff\_shell1\_NValance & 81 & max\_NeighDiff\_shell1\_SpaceGroupNumber \\
19 & mean\_NeighDiff\_shell1\_GSbandgap & 82 & var\_NeighDiff\_shell1\_Number \\
20 & max\_MeanBondLength & 83 & min\_NeighDiff\_shell1\_NpUnfilled \\
21 & max\_NeighDiff\_shell1\_Row & 84 & max\_BondLengthVariation \\
22 & range\_NeighDiff\_shell1\_NfValence & 85 & max\_NeighDiff\_shell1\_NUnfilled \\
23 & min\_NeighDiff\_shell1\_GSvolume\_pa & 86 & max\_NeighDiff\_shell1\_AtomicWeight \\
24 & range\_NeighDiff\_shell1\_NValance & 87 & min\_NeighDiff\_shell1\_NfValence \\
25 & min\_MeanBondLength & 88 & var\_NeighDiff\_shell1\_MeltingT \\
26 & mean\_NeighDiff\_shell1\_Number & 89 & mean\_NeighDiff\_shell1\_MendeleevNumber \\
27 & min\_BondLengthVariation & 90 & var\_NeighDiff\_shell1\_AtomicWeight \\
\hline
\end{tabular}
\label{tab9}
\end{table}

\begin{table}[H]
\centering
\begin{tabular}{|c|l|c|l|}
\hline
Sr. No. & Feature Name & Sr. No. & Feature Name \\
\hline
28 & mean\_NeighDiff\_shell1\_CovalentRadius & 91 & min\_NeighDiff\_shell1\_MendeleevNumber \\
29 & var\_NeighDiff\_shell1\_GSmagmom & 92 & min\_NeighDiff\_shell1\_NsUnfilled \\
30 & range\_NeighDiff\_shell1\_GSbandgap & 93 & var\_NeighDiff\_shell1\_CovalentRadius \\
31 & mean\_NeighDiff\_shell1\_NpUnfilled & 94 & var\_NeighDiff\_shell1\_NValance \\
32 & range\_NeighDiff\_shell1\_NfUnfilled & 95 & var\_NeighDiff\_shell1\_Column \\
33 & max\_NeighDiff\_shell1\_NpValence & 96 & var\_NeighDiff\_shell1\_NdUnfilled \\
34 & mean\_NeighDiff\_shell1\_NdValence & 97 & min\_NeighDiff\_shell1\_GSmagmom \\
35 & mean\_NeighDiff\_shell1\_NsValence & 98 & min\_NeighDiff\_shell1\_GSbandgap \\
36 & range\_NeighDiff\_shell1\_Number & 99 & MaxPackingEfficiency \\
37 & range\_NeighDiff\_shell1\_CovalentRadius & 100 & range\_NeighDiff\_shell1\_MeltingT \\
38 & mean\_NeighDiff\_shell1\_GSvolume\_pa & 101 & mean\_NeighDiff\_shell1\_Electronegativity \\
39 & var\_NeighDiff\_shell1\_NsValence & 102 & min\_EffectiveCoordination \\
40 & max\_NeighDiff\_shell1\_Number & 103 & mean\_NeighDiff\_shell1\_NsUnfilled \\
41 & min\_NeighDiff\_shell1\_Number & 104 & range\_NeighDiff\_shell1\_NsValence \\
42 & var\_NeighDiff\_shell1\_NsUnfilled & 105 & min\_NeighDiff\_shell1\_NdUnfilled \\
43 & var\_NeighDiff\_shell1\_NfValence & 106 & var\_NeighDiff\_shell1\_Row \\
44 & max\_NeighDiff\_shell1\_MeltingT & 107 & min\_NeighDiff\_shell1\_NsValence \\
45 & var\_NeighDiff\_shell1\_NdValence & 108 & min\_NeighDiff\_shell1\_Row \\
46 & min\_NeighDiff\_shell1\_NdValence & 109 & range\_NeighDiff\_shell1\_NdUnfilled \\
47 & var\_NeighDiff\_shell1\_Electronegativity & 110 & mean\_NeighDiff\_shell1\_NpValence \\
48 & mean\_NeighDiff\_shell1\_MeltingT & 111 & var\_NeighDiff\_shell1\_NpUnfilled \\
49 & min\_NeighDiff\_shell1\_Electronegativity & 112 & min\_NeighDiff\_shell1\_AtomicWeight \\
50 & max\_NeighDiff\_shell1\_NpUnfilled & 113 & max\_EffectiveCoordination \\
51 & range\_NeighDiff\_shell1\_NdValence & 114 & min\_NeighDiff\_shell1\_Column \\
52 & var\_NeighDiff\_shell1\_GSvolume\_pa & 115 & range\_NeighDiff\_shell1\_NpUnfilled \\
53 & mean\_EffectiveCoordination & 116 & var\_NeighDiff\_shell1\_NUnfilled \\
54 & range\_NeighDiff\_shell1\_GSmagmom & 117 & max\_NeighDiff\_shell1\_NdValence \\
55 & max\_NeighDiff\_shell1\_NfValence & 118 & range\_NeighDiff\_shell1\_Column \\
56 & var\_MeanBondLength & 119 & mean\_NeighDiff\_shell1\_NUnfilled \\

\hline
\end{tabular}
\end{table}

\begin{table}[htbp]
\centering
\begin{tabular}{|c|l|c|l|}
\hline
Sr. No. & Feature Name & Sr. No. & Feature Name \\
\hline
57 & mean\_NeighDiff\_shell1\_GSmagmom & 120 & var\_NeighDiff\_shell1\_MendeleevNumber \\
58 & var\_NeighDiff\_shell1\_NpValence & 121 & max\_NeighDiff\_shell1\_Electronegativity \\
59 & range\_NeighDiff\_shell1\_GSvolume\_pa & 122 & max\_NeighDiff\_shell1\_NsUnfilled \\
60 & mean\_BondLengthVariation & 123 & max\_NeighDiff\_shell1\_CovalentRadius \\
61 & max\_NeighDiff\_shell1\_NfUnfilled & 124 & range\_NeighDiff\_shell1\_MendeleevNumber \\
62 & mean\_NeighDiff\_shell1\_Column & 125 & mean\_NeighDiff\_shell1\_NValance \\
63 & var\_CellVolume & 126 & range\_NeighDiff\_shell1\_NUnfilled \\
\hline
\end{tabular}
\end{table}

\medskip

\begin{figure}[b]
\begin{centering}
\includegraphics[scale=0.53]{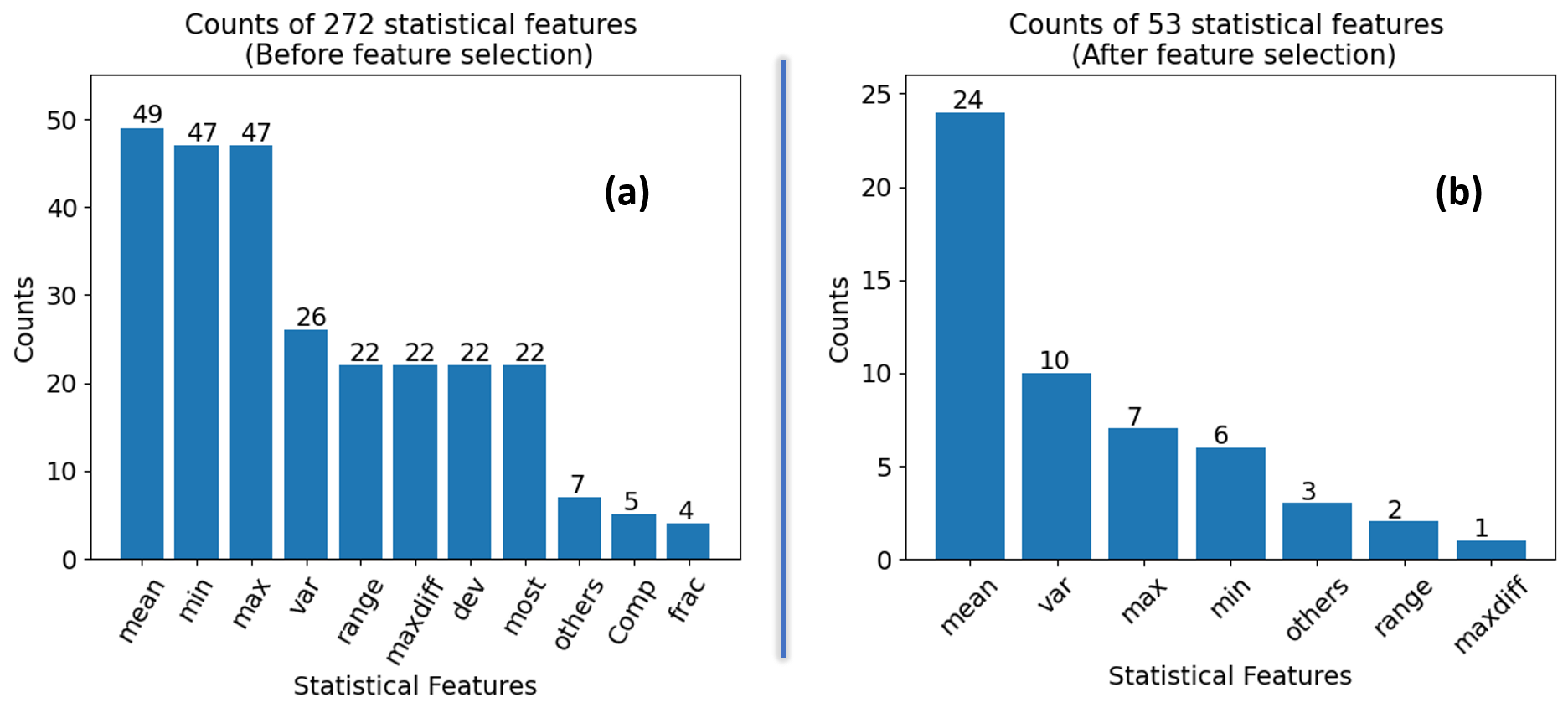}
\par\end{centering}
\caption{\protect\label{figures:figure_S2.jpg} Distribution of Statistical features (a) before feature selection (\textcolor{black}{271 Features}) (b) after feature selection (\textcolor{black}{53 Features})}
\end{figure}

\begin{table}[H]
\centering
\caption{Selected features for model training and their abbreviations used for Pearson Correlation analysis plot.}
\begin{tabular}{|l|l||l|l|}
\hline
\textbf{Feature Name} & \textbf{Abbr.} & \textbf{Feature Name} & \textbf{Abbr.} \\
\hline
Temperature & F1 & var\_NeighDiff\_shell1\_GSvolume\_pa & F28 \\
mean\_EffectiveCoordination & F2 & mean\_NeighDiff\_shell1\_GSbandgap & F29 \\
var\_EffectiveCoordination & F3 & range\_NeighDiff\_shell1\_GSbandgap & F30 \\
mean\_NeighDiff\_shell1\_Number & F4 & mean\_NeighDiff\_shell1\_SpaceGroupNumber & F31 \\
var\_NeighDiff\_shell1\_Number & F5 & var\_NeighDiff\_shell1\_SpaceGroupNumber & F32 \\
mean\_NeighDiff\_shell1\_MendeleevNumber & F6 & NComp & F33 \\
var\_NeighDiff\_shell1\_MendeleevNumber & F7 & mean\_Number & F34 \\
mean\_NeighDiff\_shell1\_MeltingT & F8 & max\_MendeleevNumber & F35 \\
var\_NeighDiff\_shell1\_MeltingT & F9 & mean\_MeltingT & F36 \\
mean\_NeighDiff\_shell1\_CovalentRadius & F10 & min\_MeltingT & F37 \\
var\_NeighDiff\_shell1\_CovalentRadius & F11 & mean\_CovalentRadius & F38 \\
max\_NeighDiff\_shell1\_Electronegativity & F12 & mean\_NpValence & F39 \\
max\_NeighDiff\_shell1\_NsValence & F13 & min\_NpValence & F40 \\
mean\_NeighDiff\_shell1\_NpValence & F14 & mean\_NdValence & F41 \\
mean\_NeighDiff\_shell1\_NdValence & F15 & max\_NdValence & F42 \\
var\_NeighDiff\_shell1\_NdValence & F16 & min\_NdValence & F43 \\
mean\_NeighDiff\_shell1\_NfValence & F17 & min\_NfValence & F44 \\
var\_NeighDiff\_shell1\_NfValence & F18 & mean\_NpUnfilled & F45 \\
mean\_NeighDiff\_shell1\_NValance & F19 & max\_NpUnfilled & F46 \\
max\_NeighDiff\_shell1\_NsUnfilled & F20 & mean\_NUnfilled & F47 \\
mean\_NeighDiff\_shell1\_NpUnfilled & F21 & min\_NUnfilled & F48 \\
range\_NeighDiff\_shell1\_NpUnfilled & F22 & min\_GSvolume\_pa & F49 \\
mean\_NeighDiff\_shell1\_NdUnfilled & F23 & maxdiff\_GSmagmom & F50 \\
mean\_NeighDiff\_shell1\_NfUnfilled & F24 & mean\_SpaceGroupNumber & F51 \\
mean\_NeighDiff\_shell1\_NUnfilled & F25 & max\_SpaceGroupNumber & F52 \\
var\_NeighDiff\_shell1\_NUnfilled & F26 & CanFormIonic & F53 \\
mean\_NeighDiff\_shell1\_GSvolume\_pa & F27 &  &  \\
\hline
\end{tabular}
\label{selected_feat_table}
\end{table}

\begin{table}[htbp]
\centering
\caption{Global average feature importance scores of Extra Trees Regressor model.}
\small
\begin{tabular}{lll}
\toprule
\textbf{Sr.No.} & \textbf{Feature Name} & \textbf{Feature Importance Score} \\
\midrule
1 & Temperature & \(1.35 \times 10^{-1}\) \\
2 & mean\_NpUnfilled & \(1.31 \times 10^{-1}\) \\
3 & min\_NUnfilled & \(1.06 \times 10^{-1}\) \\
4 & min\_GSvolume\_pa & \(7.03 \times 10^{-2}\) \\
5 & mean\_EffectiveCoordination & \(6.55 \times 10^{-2}\) \\
6 & mean\_CovalentRadius & \(6.28 \times 10^{-2}\) \\
7 & mean\_NUnfilled & \(3.16 \times 10^{-2}\) \\
8 & mean\_NeighDiff\_shell1\_MeltingT & \(3.01 \times 10^{-2}\) \\
9 & mean\_Number & \(2.32 \times 10^{-2}\) \\
10 & range\_NeighDiff\_shell1\_NpUnfilled & \(1.89 \times 10^{-2}\) \\
11 & mean\_MeltingT & \(1.87 \times 10^{-2}\) \\
12 & max\_NpUnfilled & \(1.71 \times 10^{-2}\) \\
13 & NComp & \(1.53 \times 10^{-2}\) \\
14 & mean\_NpValence & \(1.53 \times 10^{-2}\) \\
15 & mean\_NeighDiff\_shell1\_GSvolume\_pa & \(1.51 \times 10^{-2}\) \\
16 & var\_NeighDiff\_shell1\_NUnfilled & \(1.44 \times 10^{-2}\) \\
17 & min\_NdValence & \(1.38 \times 10^{-2}\) \\
18 & var\_EffectiveCoordination & \(1.32 \times 10^{-2}\) \\
19 & max\_NeighDiff\_shell1\_Electronegativity & \(1.25 \times 10^{-2}\) \\
20 & var\_NeighDiff\_shell1\_MendeleevNumber & \(1.22 \times 10^{-2}\) \\
21 & min\_NpValence & \(1.18 \times 10^{-2}\) \\
22 & mean\_NeighDiff\_shell1\_NfValence & \(1.13 \times 10^{-2}\) \\
23 & var\_NeighDiff\_shell1\_GSvolume\_pa & \(1.13 \times 10^{-2}\) \\
24 & var\_NeighDiff\_shell1\_CovalentRadius & \(1.12 \times 10^{-2}\) \\
25 & max\_MendeleevNumber & \(1.07 \times 10^{-2}\) \\
26 & mean\_NeighDiff\_shell1\_NpUnfilled & \(9.11 \times 10^{-3}\) \\
27 & mean\_NeighDiff\_shell1\_MendeleevNumber & \(8.37 \times 10^{-3}\) \\
\bottomrule
\end{tabular}
\label{feat_score_table}
\end{table}

\begin{table}[htbp]
\centering
\small
\begin{tabular}{lll}
\toprule
\textbf{Sr.No.} & \textbf{Feature Name} & \textbf{Feature Importance Score} \\
\midrule
28 & var\_NeighDiff\_shell1\_MeltingT & \(7.36 \times 10^{-3}\) \\
29 & mean\_NdValence & \(7.33 \times 10^{-3}\) \\
30 & max\_NeighDiff\_shell1\_NsValence & \(6.54 \times 10^{-3}\) \\
31 & mean\_NeighDiff\_shell1\_CovalentRadius & \(6.43 \times 10^{-3}\) \\
32 & mean\_NeighDiff\_shell1\_NdUnfilled & \(5.98 \times 10^{-3}\) \\
33 & mean\_NeighDiff\_shell1\_NpValence & \(5.94 \times 10^{-3}\) \\
34 & var\_NeighDiff\_shell1\_Number & \(5.15 \times 10^{-3}\) \\
35 & range\_NeighDiff\_shell1\_GSbandgap & \(4.68 \times 10^{-3}\) \\
36 & CanFormIonic & \(4.64 \times 10^{-3}\) \\
37 & mean\_NeighDiff\_shell1\_NUnfilled & \(4.49 \times 10^{-3}\) \\
38 & max\_SpaceGroupNumber & \(4.46 \times 10^{-3}\) \\
39 & var\_NeighDiff\_shell1\_SpaceGroupNumber & \(4.45 \times 10^{-3}\) \\
40 & mean\_NeighDiff\_shell1\_Number & \(4.37 \times 10^{-3}\) \\
41 & var\_NeighDiff\_shell1\_NdValence & \(4.18 \times 10^{-3}\) \\
42 & mean\_NeighDiff\_shell1\_NdValence & \(4.05 \times 10^{-3}\) \\
43 & min\_MeltingT & \(4.01 \times 10^{-3}\) \\
44 & mean\_SpaceGroupNumber & \(3.80 \times 10^{-3}\) \\
45 & mean\_NeighDiff\_shell1\_NValance & \(3.58 \times 10^{-3}\) \\
46 & mean\_NeighDiff\_shell1\_GSbandgap & \(3.45 \times 10^{-3}\) \\
47 & mean\_NeighDiff\_shell1\_SpaceGroupNumber & \(3.41 \times 10^{-3}\) \\
48 & max\_NeighDiff\_shell1\_NsUnfilled & \(2.03 \times 10^{-3}\) \\
49 & max\_NdValence & \(1.68 \times 10^{-3}\) \\
50 & var\_NeighDiff\_shell1\_NfValence & \(1.59 \times 10^{-3}\) \\
51 & mean\_NeighDiff\_shell1\_NfUnfilled & \(9.51 \times 10^{-4}\) \\
52 & maxdiff\_GSmagmom & \(9.43 \times 10^{-4}\) \\
53 & min\_NfValence & \(3.36 \times 10^{-5}\) \\
\bottomrule
\end{tabular}
\end{table}

\section{Machine Learning Models:}
In this study, a variety of supervised machine learning (ML) models were employed to train and validate the temperature-dependent lattice thermal conductivity ($\kappa_L$) of 150 compounds using their corresponding feature vectors as input. All ML models were implemented using the scikit-learn\citep{sklearn} python library and trained using default hyperparameters.
The regression models evaluated in this work include:
\tabularnewline
\textbf{(1) Decision Tree:}
A non-parametric, tree-based algorithm that recursively splits the dataset based on feature values, resulting in a hierarchical decision structure. Although simple and interpretable, it is often susceptible to overfitting but serves as a foundational model for many ensemble methods\citep{breiman1984classification}.
\tabularnewline
\textbf{(2) Random Forest:}
An ensemble learning method that builds multiple decision trees on randomly sampled subsets of the data and features, and averages their outputs to reduce variance and improve generalization \citep{breiman2001randomforest}.
\tabularnewline
\textbf{(3) Extreme Gradient Boosting (XGBoost):}
An optimized implementation of gradient boosting that incorporates regularization, pruning, and parallelization for efficient, accurate, and scalable learning\citep{chen2016xgboost}.
\tabularnewline
\textbf{ (4) Gradient Boosting Regressor:}
A sequential ensemble model where each new tree corrects the prediction errors of the previous trees by optimizing a loss function. It is known for strong predictive performance but typically requires careful tuning of hyperparameters\citep{friedman2001gradientboosting}.
\tabularnewline
\textbf{(5) AdaBoost Regressor (with Decision Trees):}
Adaptive Boosting combines multiple weak learners (decision trees) by reweighting the training instances, focusing more on those that were previously misclassified, to iteratively improve prediction accuracy\citep{freund1997boosting}.
\tabularnewline
\textbf{(6) Bagging Regressor (with Decision Trees):}
Bootstrap aggregation (bagging) involves training multiple decision trees on different random subsets of the training data and aggregating their outputs, reducing model variance and improving robustness\citep{breiman1996bagging}.
\tabularnewline
\textbf{ (7) Extra Trees Regressor:}
A variant of the Random Forest model, Extra Trees introduces additional randomness in selecting the features and threshold at each split, often resulting in a more generalized and diverse model\citep{geurts2006extratrees}.

\begin{figure}[h]
\begin{centering}
\includegraphics[scale=0.60]{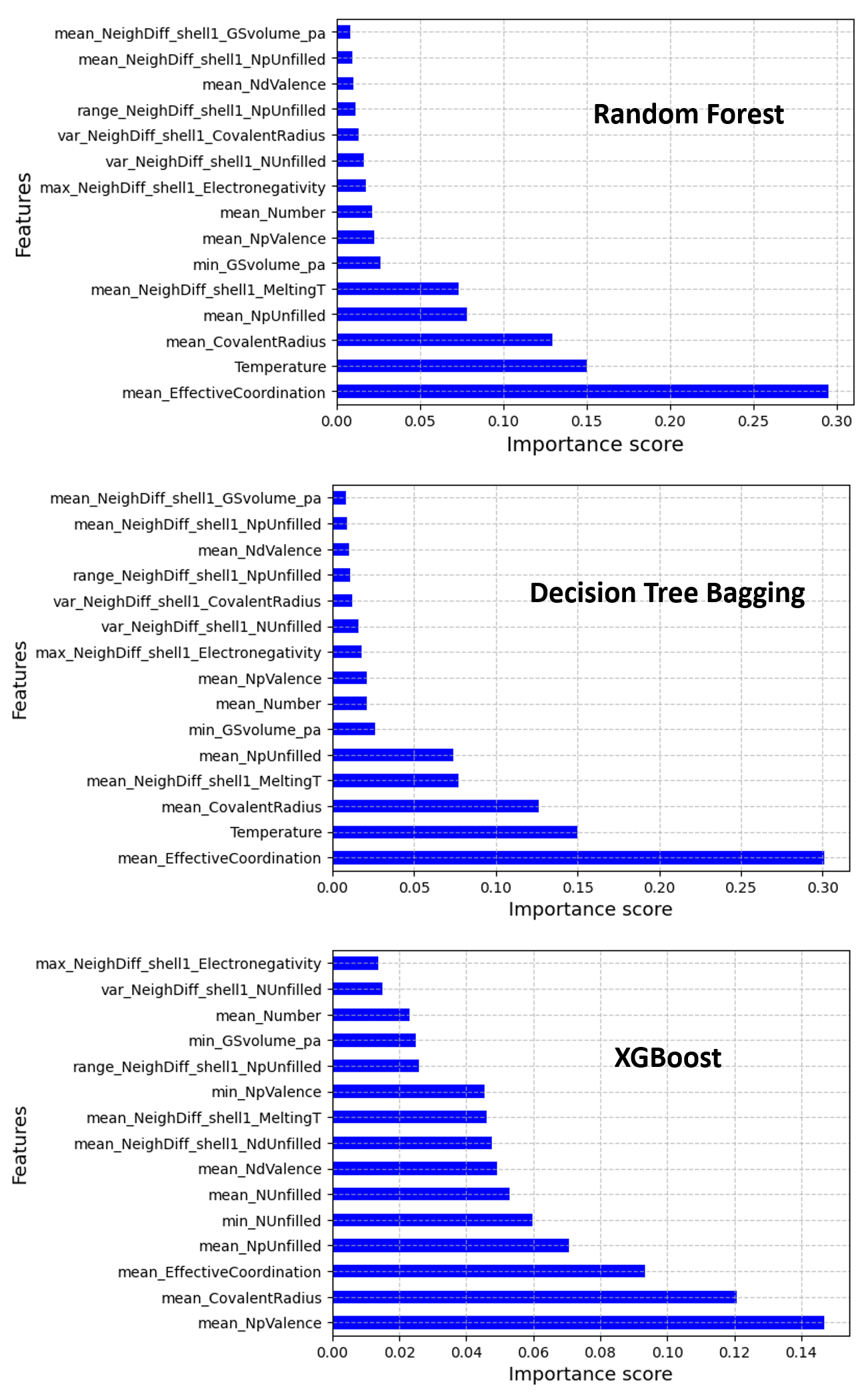}
\par\end{centering}
\caption{\protect\label{figures:figure_S3.png} Feature importance scores of Random Forest, Decision Tree Bagging and XGBoost models.}
\end{figure}

\begin{figure}[h]
\begin{centering}
\includegraphics[scale=0.60]{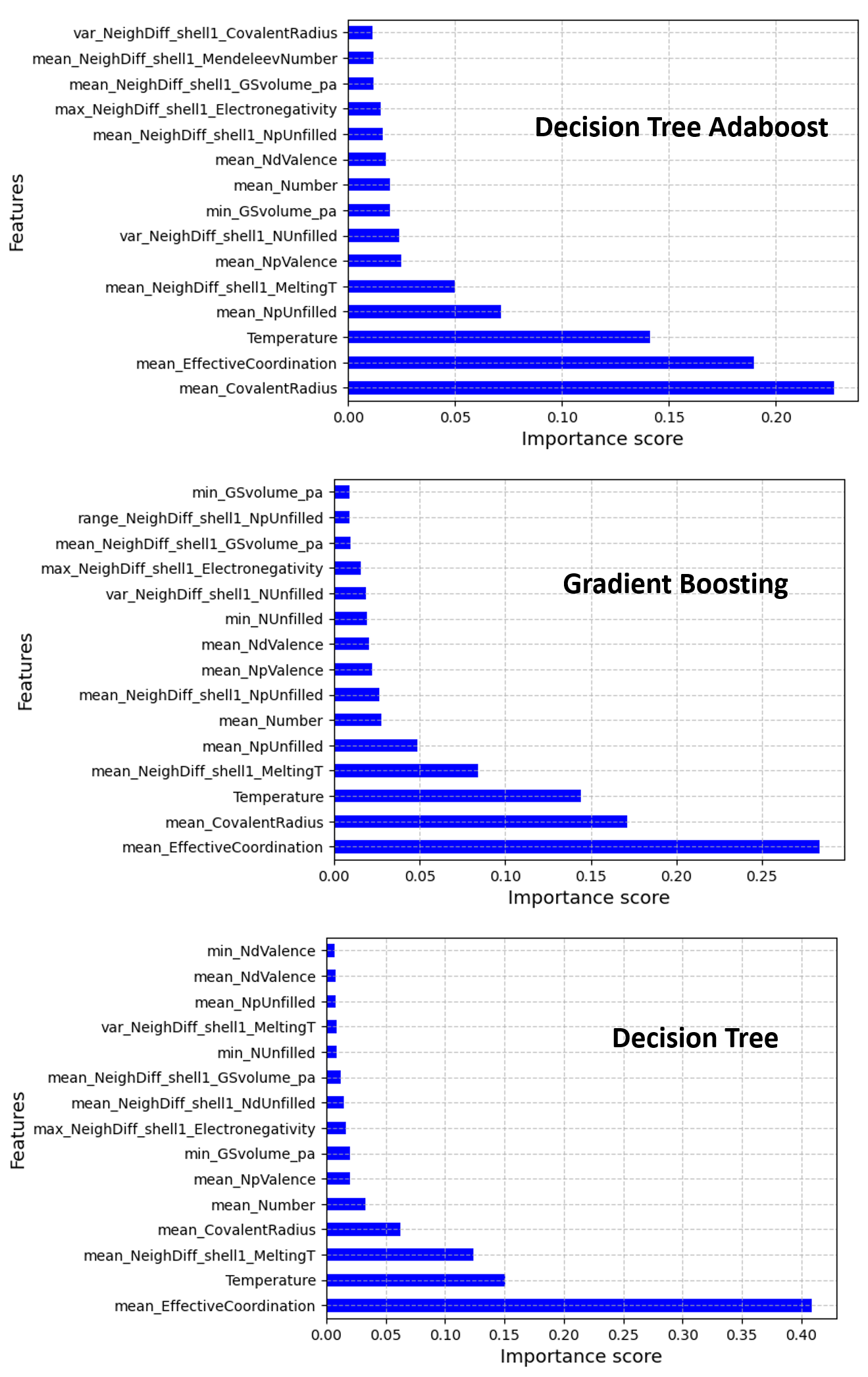}
\par\end{centering}
\caption{\protect\label{figures:figure_S4.png} Feature importance scores of Decision Tree Adaboost, Gradient Boosting and Decision Tree models.}
\end{figure}

\begin{figure}[h]
\begin{centering}
\includegraphics[scale=0.60]{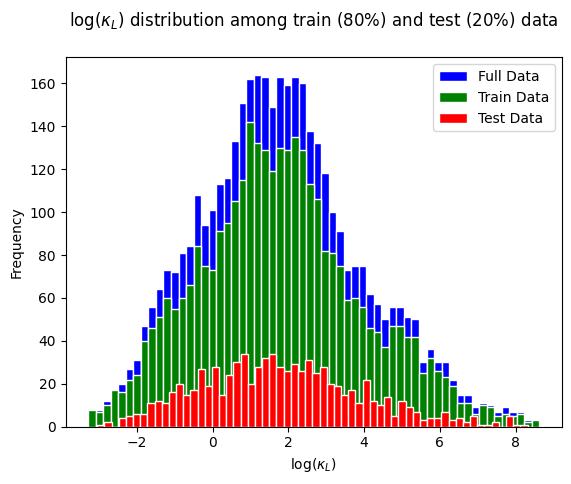}
\par\end{centering}
\caption{\protect\label{figures:figure_S5.png} Train-Test data distribution of log($\kappa_L$) for model training and testing.}
\end{figure}

\FloatBarrier
\bibliographystyle{unsrt}
\bibliography{Ref}